\documentclass[a4paper,11pt]{article}
\pdfoutput=1 

\usepackage{jheppub} 

\usepackage[T1]{fontenc} 

\usepackage{amsmath,amssymb}
\usepackage{tikz}
\usetikzlibrary{arrows.meta} 
\usepackage{hyperref} 
\usepackage{cleveref} 
\usetikzlibrary{fadings} 

\newcommand{\spacetimestrip}{%
    \begin{tikzpicture}[baseline=(current bounding box.center), scale=1.5]

        \draw[thick] (-4, 0) -- (0, 0);       
        \draw[thick] (-4, 0.6) -- (0, 0.6);   
        
        \draw[ultra thick, blue] (0, 0) -- (0, 0.6) 
            node[midway, right=3pt, black, align=left] {$\tau = 0$};

        \node[left] at (-4, 0) {$\sigma = 0$};
        \node[left] at (-4, 0.6) {$\sigma = \pi$};
         \draw[thick,->] (-5.2, 0.3) -- (-6, 0.3);       
        \node[left] at (-6, 0.3) {$\tau = -\infty$};

    \end{tikzpicture}%
}
\newcommand{\integration}{%
    \begin{tikzpicture}[baseline=(current bounding box.center), scale=1.5]

        \draw[thick,dashed] (-4, 0.2) -- (0, 0);       
         \draw[thick] (-3.6, -0.2) -- (0, 0);       
        \draw[thick] (-3.8, 0.6) -- (0, 0.6);   
\draw[black, thick] (-3.6, -0.2) .. controls (-3.8,.78) .. (-4, 0.2);
        
        \draw[ultra thick, blue] (0, 0) -- (0, 0.6) 
            node[midway, right=3pt, black, align=left] {$\tau = 0$};

        \node[left] at (-4, -.2) {$\sigma = 0$};
        \node[left] at (-4, 0.2) {$\sigma = \pi$};

    \end{tikzpicture}%
}
\newcommand{\starproduct}{%
    \begin{tikzpicture}[baseline=(current bounding box.center), scale=1.5]
        
\draw[black, thick] (-2,2) .. controls (-.3,.75) .. (2,1);
     \node[left] at (-2,1.5) {$\Psi_1$};
          \node[left] at (2, .5) {$\Psi_2$};

\draw[black, thick] (.15,-.2) .. controls (.3,.-.1) .. (2,0);

\draw[black, thick] (-2.5,1.2) .. controls (-1,.2) .. (-.9,0);

\draw[blue, dotted, very thick] (-.3,.45) .. controls (-.1,.45) .. (0,.9);

\draw[blue, very thick] (-.3,.45) .. controls (-.2,0.1) .. (0.15,-.2);

\draw[blue, very thick] (-.3,.45) .. controls (-.5,0.1) .. (-.9,0);

    \filldraw[red] (-.3,.45) circle (1.5 pt);
    \end{tikzpicture}%
}

\newcommand{\wittenvertex}{%
    \begin{tikzpicture}[baseline=(current bounding box.center), scale=1.5]
        
\draw[black, thick] (-2,2) .. controls (-.3,.75) .. (2,1);
   \node[left] at (-2,1.5) {$\Psi_1$};
          \node[left] at (2, .5) {$\Psi_2$};
          \node[left] at (-.45, -1.4) {$\Psi_3$};

\draw[black, thick] (.15,-.2) .. controls (.3,.-.1) .. (2,0);

\draw[black, thick] (-2.5,1.2) .. controls (-1,.2) .. (-.9,0);

\draw[blue, dotted, very thick] (-.3,.45) .. controls (-.1,.45) .. (0,.9);

\draw[blue, dotted, very thick] (-.3,.45) .. controls (-.2,0.1) .. (0.15,-.2);

\draw[blue, dotted,  very thick] (-.3,.45) .. controls (-.5,0.1) .. (-.9,0);

\draw[black, thick] (.15,-.2) -- (-.25,-1.75);

\draw[black, thick] (-.9,0) -- (-1.25,-1.5);

    \filldraw[red] (-.3,.45) circle (1.5 pt);
    \end{tikzpicture}%
}

\newcommand{\tpointloop}{%
\begin{tikzpicture}[line join=round, line cap=round, >=stealth, scale=1]

    \def\R{1.2}      
    \def\H{3.5}      
    \def\ry{0.35}    
    \def\slitH{1}  

    \draw[ thick] (\R, \H) arc (0:180:\R\space and \ry);
    \draw[dashed, thick, gray!70] (\R, 0) arc (0:180:\R\space and \ry);
    \draw[ thick, blue!70,<->] (.8, 1.8) arc (310:250:\R\space and \ry);
\draw[dashed, thick, blue!70,<->] (.9, 1.85) arc (-40:250:\R\space and \ry);
    \draw[thick] (-\R, 0) -- (-\R, \H);
    \draw[thick] (\R, 0) -- (\R, \H);

    \draw[thick] (-1.2, 3.5) arc (180:300:\R\space and \ry) -- (.8,1.75);
    \draw[dashed, thick] (1, 3.35) -- (.8, 1.75);
    \draw[thick] (1, 3.35) arc (320:360: .9 \space and .25) ;

    \draw[dashed, thick] (-\R, 0) arc (180:240:\R\space and \ry) -- (-.4, 1.75);
    \draw[thick] (-.2, -.35) -- (-.4, 1.75);
    \draw[thick] (-.2, -.35) arc (260:360: 1.2 \space and .35) ;

    \draw[thick] (-2.2, -.4) -- (-2, 1.65);
    \draw[thick] (-1.8, -.45) -- (-2, 1.65);
      \draw[thick] (-2, 1.65) -- (-.4, 1.75);
      \draw[ thick] (-2, 1.65) -- (-.4, 1.75);
      \draw[thick] (-1.8, -.45) -- (-.2, -.35);
       \draw[dashed, thick] (-2.2, -.4) -- (-.625, -.3);
         \draw[thick] (1, 3.35) -- (2.5, 3.35);
    \draw[thick] (.6, 3.2) -- (2.3, 3.2);
      \draw[thick] (.8, 1.75) -- (2.4, 1.75);
      \draw[thick] (2.3, 3.2) -- (2.4, 1.75);
      \draw[thick] (2.4, 1.75) -- (2.5, 3.35);
  
  \node[below] at (0, 1.75) {$T_1$};
   \node[above] at (0, 2.4) {$T_2$};
    \node[below] at (4,1.75) {$T_1=T_2=0$};
      \node[below] at (-2.5, 0) {$\Psi_1$};
   \node[above] at (2.8, 2.4) {$\Psi_2$};
     \node[below] at (5.5, 0) {$\Psi_1$};
   \node[above] at (9.8, 2.4) {$\Psi_2$};
    
    \draw[->,, thick] (3, 1.75) -- (5, 1.75);

    \draw[thick] (5.8, -.4) -- (6, 1.65);
    \draw[thick] (6.2, -.45) -- (6, 1.65);
      \draw[thick] (6, 1.65) -- (7.6, 1.75);
      \draw[thick] (6, 1.65) -- (7.6, 1.75);
      \draw[thick] (6.2, -.45) -- (7.6, -.35);
       \draw[dashed, thick] (5.8, -.4) -- (7.6, -.35);
       \draw[thick] (7.6, 1.75) -- (7.6, -.35);
       
         \draw[thick] (7.6, 3.2) -- (9.5, 3.35);
    \draw[thick] (7.6, 3.2) -- (9.3, 3.2);
      \draw[thick] (7.6, 1.75) -- (9.4, 1.75);
      \draw[thick] (9.3, 3.2) -- (9.4, 1.75);
      \draw[thick] (9.4, 1.75) -- (9.5, 3.35);
 \draw[thick] (7.6, 1.75) -- (7.6, 3.2);
\end{tikzpicture}

}

\newcommand{\opentclosed}{%
    \begin{tikzpicture}[baseline=(current bounding box.center), scale=1.5]

        \draw[thick,dashed] (-4, 0.2) -- (0, 0);       
         \draw[thick] (-3.6, -0.2) -- (0, 0);       
        \draw[thick] (-3.8, 0.6) -- (0, 0.6);   
\draw[black, thick] (-3.6, -0.2) .. controls (-3.8,.78) .. (-4, 0.2);
        
        \draw[ultra thick, blue] (0, 0) -- (0, 0.6) 
            node[midway, right=3pt, black, align=left] {$\tau = 0$};

        \node[left] at (-3.75, 0) {$\Psi$};
    \filldraw[red] (0, 0.6) circle (1.5 pt);
 \node[above] at (0, 0.6) {$V$};
    \end{tikzpicture}%
}

\newcommand{\wedgen}{%
    \begin{tikzpicture}[baseline=(current bounding box.center), scale=1.5]
    
    \draw[blue!80!black, thick] (-1,0) arc(180:90:1);
    \draw[green!80!black, thick] (1,0) arc(0:90:1);
    \draw[red!80!black, thick] (-1,0) -- (1,0);
    
    \draw[->] (-1.5,0) -- (1.5,0) node[right, scale=0.8] {$\text{Re}(z)$};
    \draw[->] (0,-0.2) -- (0,1.4) node[above, scale=0.8] {$\text{Im}(z)$};
    
    \node[below, scale=0.8] at (1,0) {$1$};
    \node[below, scale=0.8] at (-1,0) {$-1$};
    \node[above left, scale=0.8] at (0,1) {$i$};
    \node[below left, scale=0.7] at (0,0) {$0$};

    \draw[->, thick] (1.8,0.5) -- (2.8,0.5) 
        node[midway, above, scale=0.75] {$f_n(z) = \frac{2}{n}\arctan z$};

    \begin{scope}[shift={(4.5,0)}]
        \def\b{0.8} 
        
        \draw[blue!80!black, thick] (-\b,1.5) -- (-\b,0);
        \draw[green!80!black, thick] (\b,1.5) -- (\b,0);
        \draw[red!80!black, thick] (-\b,0) -- (\b,0);
        
        \draw[->] (-1.5,0) -- (1.5,0) node[right, scale=0.8] {$\text{Re}(w)$};
        \draw[->] (0,-0.2) -- (0,1.7) node[above, scale=0.8] {$\text{Im}(w)$};
        
        \node[below, scale=0.75] at (\b,0) {$\frac{n}{2}$};
        \node[below, scale=0.75] at (-\b,0) {$-\frac{n}{2}$};
        \node[below left, scale=0.7] at (0,0) {$0$};
    \end{scope}

                \end{tikzpicture}%
}

\newcommand{\potplot}{%
    \begin{tikzpicture}[baseline=(current bounding box.center), scale=1.6]

    \draw[->] (-1.5,0) -- (2.5,0) node[right] {$\phi$};
    \draw[->] (0,-1) -- (0,1.5) node[above] {$V(\phi)$};

    \draw[blue, thick, samples=100, domain=0:1.85] plot (\x, {-1.5*\x*\x + \x*\x*\x});

    \fill[red] (0,0) circle (2pt);
    \node[above right, scale=0.8, text=red] at (0,0.1) {Perturbative vacuum};

    \fill[green!60!black] (1,-0.5) circle (2pt);
    \node[below right, scale=0.8, text=green!60!black] at (1,-0.5) {Tachyon vacuum};

    \end{tikzpicture}%
}

\newcommand{\innerp}{%
    \begin{tikzpicture}[baseline=(current bounding box.center), scale=1.6]

    \draw[red, thick] (0,0) -- (0,2);
    \draw[green!60!black, thick] (1,0) -- (1,2);
    \draw (0,0) -- (1,0);
    \node[below, scale=0.8] at (0.5,0) { $\Omega^{t_1}$};

    \begin{scope}[shift={(1.5,0)}]
        \draw[green!60!black, thick] (0,0) -- (0,2);
        \draw[red, thick] (0.8,0) -- (0.8,2);
        \draw (0,0) -- (0.8,0);
        \node[below, scale=0.8] at (0.4,0) { $\Omega^{t_2}$};
    \end{scope}

    \draw[->, thick, gray] (2.6, 1) -- (3.2, 1) node[midway, above, scale=0.8, text=black] {$\int \Omega^{t_1} \ast \Omega^{t_2}$};

    \begin{scope}[shift={(4.5,0)}]
        
        \draw (0,0) ellipse (0.6cm and 0.2cm);
         \draw (0,2) ellipse (0.6cm and 0.2cm);
        \draw[gray!50] (-0.6,0) -- (-0.6,2);
        \draw[gray!50] (0.6,0) -- (0.6,2);
        
        \draw[red, thick] (-0.3, -0.17) -- (-0.3, 1.83);
        \draw[green!60!black, thick] (0.3, -0.17) -- (0.3, 1.83);
        
        \node[below, scale=0.8] at (0,-0.2) { $t_1+t_2$};
    \end{scope}

    \end{tikzpicture}%
}

\newcommand{\prop}{%
    \begin{tikzpicture}[baseline=(current bounding box.center), scale=1.3]

    \begin{scope}[xshift=0cm, yshift=1cm]
        \draw (-1.5, 0) -- (0, 0); 
        \node[left] at ( -1.5,0) { $\left(k_i,\beta_i\right)$};
        \node[right] at (0, 0) { $\left(k_j,\beta_j\right)$};
    \end{scope}

    \draw[->, line width=1pt] (1.2, 1) -- (1.8, 1); 
      \node[right] at ( 2,1) { $\int_{0}^{\infty} dt ~e^{-t}$};

    \begin{scope}[xshift=5cm, yshift=0cm]
        \draw (0, 2) ellipse (1.2cm and 0.3cm);       
        \node[above] at (-.1,1.75){$t$};
         \node[above] at (-1.2,2.1){$1$};
          \node[above] at (1.2,2.1){$1$};
        \draw (-1.2, 0) arc (180:360:1.2cm and 0.3cm); 
        \draw [dashed](1.2, 0) arc (0:180:1.2cm and 0.3cm); 
        \filldraw[black] ( -1.2,0) circle (1.5 pt);
        \filldraw[black] ( 1.2,0) circle (1.5 pt);
         \node[left] at ( -1.2,0) { $\mathcal{V}_{k_i}^{\beta_i}$};
        \node[right] at (1.2, 0) { $\mathcal{V}_{k_j}^{\beta_j}$};
        \draw (-1.2, 0) -- (-1.2, 2);                  
        \draw (1.2, 0)  -- (1.2, 2);                   

        \draw[very thick] (-0.8, -0.22) -- (-0.8, 1.78)node[pos=0.6] {$\quad B$};
        \draw (0.8, -0.22)  -- (0.8, 1.78);  
        \draw[dashed, gray!60] (0, 0.3) -- (0, 2.3);   
    \end{scope}

    \end{tikzpicture}%
}

\newcommand{\cubicv}{%
    \begin{tikzpicture}[baseline=(current bounding box.center), scale=1.3]

    \begin{scope}[xshift=0cm, yshift=1cm]
        \fill (0,0) circle (2.5pt); 
        
        \draw (0,0) -- (0, 1.3) node[above] {$\left(k_l,\beta_l\right)$};    
        \draw (0,0) -- (-1.1, -0.65) node[below] {$\left(k_i,\beta_i\right)$}; ; 
        \draw (0,0) -- (1.1, -0.65) node[below] {$\left(k_j,\beta_j\right)$}; ;  
    \end{scope}

    \draw[->, line width=1pt] (1.8, 1) -- (3.2, 1);

    \begin{scope}[xshift=5cm, yshift=0cm]
        \draw (0, 2) ellipse (1.2cm and 0.3cm);       
         \node[above] at (-.1,1.75){$1$};
         \node[above] at (-1.2,2.1){$1$};
          \node[above] at (1.2,2.1){$1$};
                \filldraw[black] ( -1.2,0) circle (1.5 pt);
        \filldraw[black] ( 1.2,0) circle (1.5 pt);
        \filldraw[black] ( 0,-.3) circle (1.5 pt);
         \node[left] at ( -1.2,0) { $\mathcal{V}_{k_i}^{\beta_i}$};
        \node[right] at (1.2, 0) { $\mathcal{V}_{k_j}^{\beta_j}$};
        \node[right] at (0, -.5) { $\mathcal{V}_{k_l}^{\beta_l}$};
        \draw (-1.2, 0) arc (180:360:1.2cm and 0.3cm); 
         \draw [dashed](1.2, 0) arc (0:180:1.2cm and 0.3cm); 
        
        \draw (-1.2, 0) -- (-1.2, 2);                  
        \draw (1.2, 0)  -- (1.2, 2);                   

        \draw (-0.8, -0.22) -- (-0.8, 1.78); 
        \draw (0.8, -0.22)  -- (0.8, 1.78);  
        \draw[dashed, gray!60] (0, 0.3) -- (0, 2.3);   
    \end{scope}

    \end{tikzpicture}%
}
\newcommand{\ProptVA}{%
    \begin{tikzpicture}[baseline=(current bounding box.center), scale=1]

    \draw[gray!25, line width=1pt] (-3.3, 0) -- (-1.7,0);

    \begin{scope}[xshift=-4cm, yshift=0cm]
        \draw (0, 2) ellipse (0.7cm and 0.175cm);
        \node[above] at (-0.05, 1.8){$1$}; \node[above] at (-0.7, 2.05){$1$}; \node[above] at (0.7, 2.05){$1$};
        \filldraw[black] (-0.7, 0) circle (1.5 pt); \filldraw[black] (0.7, 0) circle (1.5 pt); \filldraw[black] (0, -0.175) circle (1.5 pt);
        \node[left, font=\scriptsize, inner sep=1pt] at (-0.7, 0) { $\mathcal{V}_{k_i}^{\beta_i}$};
        \node[right, font=\scriptsize, inner sep=1pt] at (0.7, 0) { ${\mathcal{V}}_{k_j}^{\beta_j}$};
        \node[right, font=\scriptsize, inner sep=2pt] at (0, -0.35) { $\mathcal{V}_{k_l}^{\beta_l}$};
        \draw (-0.7, 0) arc (180:360:0.7cm and 0.175cm); \draw [dashed](0.7, 0) arc (0:180:0.7cm and 0.175cm);
        \draw (-0.7, 0) -- (-0.7, 2); \draw (0.7, 0)  -- (0.7, 2);
        \draw (-0.46, -0.13) -- (-0.46, 1.87); \draw [green] (0.46, -0.13)  -- (0.46, 1.87); \draw[dashed, blue] (0, 0.175) -- (0, 2.175);
    \end{scope}

    \begin{scope}[xshift=-1cm, yshift=0cm]
        \draw (0, 2) ellipse (0.7cm and 0.175cm);
        \node[above] at (-0.05, 1.8){$t$}; \node[above] at (-0.7, 2.05){$1$}; \node[above] at (0.7, 2.05){$1$};
        \filldraw[black] (-0.7, 0) circle (1.5 pt); \filldraw[black] (0.7, 0) circle (1.5 pt); \filldraw[black] (0, -0.175) circle (1.5 pt);
        \node[left, font=\scriptsize, inner sep=1pt] at (-0.7, 0) { ${\mathcal{V}^c}_{k_j}^{\beta_j}$};
        \node[right, font=\scriptsize, inner sep=1pt] at (0.7, 0) { $\mathcal{V}_{k}^{\alpha}$};
        \draw (-0.7, 0) arc (180:360:0.7cm and 0.175cm); \draw [dashed](0.7, 0) arc (0:180:0.7cm and 0.175cm);
        \draw (-0.7, 0) -- (-0.7, 2); \draw (0.7, 0)  -- (0.7, 2);
        \draw [green](-0.46, -0.13) -- (-0.46, 1.87); \draw [very thick](0.46, -0.13)  -- (0.46, 1.87)node[pos=0.6] {$\quad B$};; \draw[dashed, blue] (0, 0.175) -- (0, 2.175);
    \end{scope}

    \draw[->, line width=1pt] (.8, 1) -- (1.8, 1);
    \begin{scope}[xshift=5cm, yshift=0cm]
        \draw (0, 2) ellipse (1.2cm and 0.3cm);       
         \node[above] at (-.1,1.75){$1$};
         \node[above] at (-.1, 2.3){$1$};
         \node[above] at (-1.2,2.1){$1$};
          \node[above] at (1.2,2.1){$t$};
                \filldraw[black] ( -1.2,0) circle (1.5 pt);
        \filldraw[black] ( 1.2,0) circle (1.5 pt);
        \filldraw[black] ( 0,-.3) circle (1.5 pt);
         \node[left] at ( -1.2,0) { $\mathcal{V}_{k_i}^{\beta_i}$};
        \node[right] at (1.2, 0) { $\mathcal{V}_{k}^{\alpha}$};
        \node[right] at (0, -.5) { $\mathcal{V}_{k_l}^{\beta_l}$};
        \draw (-1.2, 0) arc (180:360:1.2cm and 0.3cm); 
         \draw [dashed](1.2, 0) arc (0:180:1.2cm and 0.3cm); 
        
        \draw (-1.2, 0) -- (-1.2, 2);                  
        \draw (1.2, 0)  -- (1.2, 2);                   

        \draw (-0.8, -0.22) -- (-0.8, 1.78); 
        \draw[green](0.8, -0.22)  -- (0.8, 1.78);  
        \draw[dashed, blue] (-.4, 0.25) -- (-.4, 2.25);   
        \draw[very thick, dashed] (.4, 0.25) -- (.4, 2.25) node[pos=0.6] {$\quad B$};   
    \end{scope}

    \end{tikzpicture}%
}

\newcommand{\ProptV}{%
    \begin{tikzpicture}[baseline=(current bounding box.center), scale=.8]

    \draw[gray!25, line width=1pt] (-5.3, 2) -- (-4.7,  0) ;
    \draw[gray!25, line width=1pt] (-5.3,-2) -- (-4, -.2);
    \draw[gray!25, line width=1pt] (-3.3, 0) -- (-1.7,0);

    \begin{scope}[xshift=-4cm, yshift=0cm]
        \draw (0, 2) ellipse (0.7cm and 0.175cm);
        \node[above] at (-0.05, 1.8){$1$}; \node[above] at (-0.7, 2.05){$1$}; \node[above] at (0.7, 2.05){$1$};
        \filldraw[black] (-0.7, 0) circle (1.5 pt); \filldraw[black] (0.7, 0) circle (1.5 pt); \filldraw[black] (0, -0.175) circle (1.5 pt);
        \draw (-0.7, 0) arc (180:360:0.7cm and 0.175cm); \draw [dashed](0.7, 0) arc (0:180:0.7cm and 0.175cm);
        \draw (-0.7, 0) -- (-0.7, 2); \draw (0.7, 0)  -- (0.7, 2);
        \draw (-0.46, -0.13) -- (-0.46, 1.87); \draw (0.46, -0.13)  -- (0.46, 1.87); \draw[dashed, gray!60] (0, 0.175) -- (0, 2.175);
    \end{scope}

    \begin{scope}[xshift=-1cm, yshift=0cm]
        \draw (0, 2) ellipse (0.7cm and 0.175cm);
        \node[above] at (-0.05, 1.8){$t_3$}; \node[above] at (-0.7, 2.05){$1$}; \node[above] at (0.7, 2.05){$1$};
        \filldraw[black] (-0.7, 0) circle (1.5 pt); \filldraw[black] (0.7, 0) circle (1.5 pt); \filldraw[black] (0, -0.175) circle (1.5 pt);
        \node[right, font=\scriptsize, inner sep=1pt] at (0.7, 0) { $\mathcal{V}_{k_3}^{\alpha_3}$};
        \draw (-0.7, 0) arc (180:360:0.7cm and 0.175cm); \draw [dashed](0.7, 0) arc (0:180:0.7cm and 0.175cm);
        \draw (-0.7, 0) -- (-0.7, 2); \draw (0.7, 0)  -- (0.7, 2);
        \draw (-0.46, -0.13) -- (-0.46, 1.87); \draw [very thick](0.46, -0.13)  -- (0.46, 1.87)node[pos=0.6] {$\quad B$}; \draw[dashed, gray!60] (0, 0.175) -- (0, 2.175);
    \end{scope}

    \begin{scope}[xshift=-6cm, yshift=2cm]
        \draw (0, 2) ellipse (0.7cm and 0.175cm);
        \node[above] at (-0.05, 1.8){$t_1$}; \node[above] at (-0.7, 2.05){$1$}; \node[above] at (0.7, 2.05){$1$};
        \filldraw[black] (-0.7, 0) circle (1.5 pt); \filldraw[black] (0.7, 0) circle (1.5 pt); 
        \node[left, font=\scriptsize, inner sep=1pt] at (-0.4, -.4) { $\mathcal{V}_{k_1}^{\alpha_1}$};
        \draw (-0.7, 0) arc (180:360:0.7cm and 0.175cm); \draw [dashed](0.7, 0) arc (0:180:0.7cm and 0.175cm);
        \draw (-0.7, 0) -- (-0.7, 2); \draw (0.7, 0)  -- (0.7, 2);
        \draw[very thick] (-0.46, -0.13) -- (-0.46, 1.87)node[pos=0.6] {$\quad B$};; \draw (0.46, -0.13)  -- (0.46, 1.87); \draw[dashed, gray!60] (0, 0.175) -- (0, 2.175);
    \end{scope}

    \begin{scope}[xshift=-6cm, yshift=-2cm]
        \draw (0, 2) ellipse (0.7cm and 0.175cm);
        \node[above] at (-0.05, 1.8){$t_2$}; \node[above] at (-0.7, 2.05){$1$}; \node[above] at (0.7, 2.05){$1$};
        \filldraw[black] (-0.7, 0) circle (1.5 pt); \filldraw[black] (0.7, 0) circle (1.5 pt); \filldraw[black] (0, -0.175) circle (1.5 pt);
        \node[left, font=\scriptsize, inner sep=1pt] at (-0.7, 0) { $\mathcal{V}_{k_2}^{\alpha_2}$};
        \draw (-0.7, 0) arc (180:360:0.7cm and 0.175cm); \draw [dashed](0.7, 0) arc (0:180:0.7cm and 0.175cm);
        \draw (-0.7, 0) -- (-0.7, 2); \draw (0.7, 0)  -- (0.7, 2);
        \draw [very thick](-0.46, -0.13) -- (-0.46, 1.87) node[pos=0.6] {$\quad B$};; \draw (0.46, -0.13)  -- (0.46, 1.87); \draw[dashed, gray!60] (0, 0.175) -- (0, 2.175);
    \end{scope}

    \draw[->, line width=1pt] (.8, 1) -- (1.8, 1);

    \begin{scope}[xshift=4cm, yshift=1cm]
        \draw (0, 1.5) ellipse (1.2cm and 0.4cm);
        \draw (-1.2, -1.5) arc (180:360:1.2cm and 0.4cm);
        \draw[dashed, gray] (-1.2, -1.5) arc (180:0:1.2cm and 0.4cm);
        
        \draw   (-1.2, -1.5) -- (-1.2, 1.5);
        \draw   (1.2, -1.5) -- (1.2, 1.5);
        
        \draw[very thick] (-0.5, -1.85) -- (-0.5, 1.15) node[pos=0.6] {$\quad B$};
        \draw[dashed, very thick] (-0.1, -1.1)  -- (-0.1, 1.9) node[pos=0.6] {$\quad B$};
        \draw[very thick] (0.5, -1.85) -- (0.5, 1.15)node[pos=0.6] {$\quad B$} ;
        
        \node[font=\footnotesize] at (-0.9, 2) {$t_1+1$};
        \filldraw[black] (-1.2, -1.5) circle (1.5 pt);
        \node[below] at  (-1.3, -1.5) {$\mathcal{V}_{k_1}^{\alpha_1}$};
        \filldraw[black] (0, -1.75) circle (1.5 pt);
        \node at  (0, -2.2) {$\mathcal{V}_{k_2}^{\alpha_2}$};
        \node[font=\footnotesize] at (0.1, 1.3) {$t_2+1$};
         \node[below] at  (1.4, -1.5) {$\mathcal{V}_{k_3}^{\alpha_3}$};
        \filldraw[black] (1.2, -1.5)  circle (1.5 pt);
        \node[font=\footnotesize] at (0.85, 2) {$t_3+1$};
    \end{scope}

    \end{tikzpicture}%
}
\newcommand{\schannG}{%
    \begin{tikzpicture}[baseline=(current bounding box.center), scale=1.05]

    \draw[gray!25, line width=1pt] (-5.3, 1.9) -- (-4.7, 0) ;
     \draw[gray!25, line width=1pt] (-3.3, 0) -- (-2.2, 0);
     \draw[gray!25, line width=1pt] (-.8, 0) -- (0.3, 0);
     \draw[gray!25, line width=1pt] (1.7, 0) -- (2.3, 1.9);
    \draw[gray!25, line width=1pt] (-5.25,-2) -- (-4, -.1);
    \draw[gray!25, line width=1pt] (1, -.2) -- (2.3,-1.95);

    \begin{scope}[xshift=-4cm, yshift=0cm]
        \draw (0, 2) ellipse (0.7cm and 0.175cm);
        \node[above] at (-0.05, 1){$1$}; \node[above] at (-0.7, 2.05){$1$}; \node[above] at (0.7, 2.05){$1$};
        \filldraw[black] (-0.7, 0) circle (1.5 pt); \filldraw[black] (0.7, 0) circle (1.5 pt); \filldraw[black] (0, -0.175) circle (1.5 pt);
        \draw (-0.7, 0) arc (180:360:0.7cm and 0.175cm); \draw [dashed](0.7, 0) arc (0:180:0.7cm and 0.175cm);
        \draw (-0.7, 0) -- (-0.7, 2); \draw (0.7, 0)  -- (0.7, 2);
        \draw (-0.46, -0.13) -- (-0.46, 1.87); \draw (0.46, -0.13)  -- (0.46, 1.87); \draw[dashed, gray!60] (0, 0.175) -- (0, 2.175);
    \end{scope}

    \begin{scope}[xshift=-1.5cm, yshift=0cm]
        \draw (0, 2) ellipse (0.7cm and 0.175cm);
        \node[above] at (-0.05, 1){$t_5$}; \node[above] at (-0.7, 2.05){$1$}; \node[above] at (0.7, 2.05){$1$};
        \filldraw[black] (-0.7, 0) circle (1.5 pt); \filldraw[black] (0.7, 0) circle (1.5 pt); 
        \draw (-0.7, 0) arc (180:360:0.7cm and 0.175cm); \draw [dashed](0.7, 0) arc (0:180:0.7cm and 0.175cm);
        \draw (-0.7, 0) -- (-0.7, 2); \draw (0.7, 0)  -- (0.7, 2);
        \draw[very thick] (-0.46, -0.13) -- (-0.46, 1.87); \draw (0.46, -0.13)  -- (0.46, 1.87); \draw[dashed, gray!60] (0, 0.175) -- (0, 2.175);
    \end{scope}

    \begin{scope}[xshift=1cm, yshift=0cm]
        \draw (0, 2) ellipse (0.7cm and 0.175cm);
        \node[above] at (-0.05, 1){$1$}; \node[above] at (-0.7, 2.05){$1$}; \node[above] at (0.7, 2.05){$1$};
        \filldraw[black] (-0.7, 0) circle (1.5 pt); \filldraw[black] (0.7, 0) circle (1.5 pt); \filldraw[black] (0, -0.175) circle (1.5 pt);
        \draw (-0.7, 0) arc (180:360:0.7cm and 0.175cm); \draw [dashed](0.7, 0) arc (0:180:0.7cm and 0.175cm);
        \draw (-0.7, 0) -- (-0.7, 2); \draw (0.7, 0)  -- (0.7, 2);
        \draw (-0.46, -0.13) -- (-0.46, 1.87); \draw (0.46, -0.13)  -- (0.46, 1.87); \draw[dashed, gray!60] (0, 0.175) -- (0, 2.175);
    \end{scope}

    \begin{scope}[xshift=-6cm, yshift=2cm]
        \draw (0, 2) ellipse (0.7cm and 0.175cm);
        \node[above] at (-0.05, 1){$t_1$}; \node[above] at (-0.7, 2.05){$1$}; \node[above] at (0.7, 2.05){$1$};
        \filldraw[black] (-0.7, 0) circle (1.5 pt); \filldraw[black] (0.7, 0) circle (1.5 pt); 
        \node[below, font=\scriptsize, inner sep=1pt] at (-0.6, -.2) { $\mathcal{V}_{k_1}^{\alpha_1}$};
        \draw (-0.7, 0) arc (180:360:0.7cm and 0.175cm); \draw [dashed](0.7, 0) arc (0:180:0.7cm and 0.175cm);
        \draw (-0.7, 0) -- (-0.7, 2); \draw (0.7, 0)  -- (0.7, 2);
        \draw[very thick](-0.46, -0.13) -- (-0.46, 1.87); \draw (0.46, -0.13)  -- (0.46, 1.87); \draw[dashed, gray!60] (0, 0.175) -- (0, 2.175);
    \end{scope}

    \begin{scope}[xshift=-6cm, yshift=-2cm]
        \draw (0, 2) ellipse (0.7cm and 0.175cm);
        \node[above] at (-0.05, 1){$t_2$}; \node[above] at (-0.7, 2.05){$1$}; \node[above] at (0.7, 2.05){$1$};
        \filldraw[black] (-0.7, 0) circle (1.5 pt); \filldraw[black] (0.7, 0) circle (1.5 pt); \filldraw[black] (0, -0.175) circle (1.5 pt);
        \node[below, font=\scriptsize, inner sep=1pt] at (-0.6, -.2) { $\mathcal{V}_{k_2}^{\alpha_2}$};
        \draw (-0.7, 0) arc (180:360:0.7cm and 0.175cm); \draw [dashed](0.7, 0) arc (0:180:0.7cm and 0.175cm);
        \draw (-0.7, 0) -- (-0.7, 2); \draw (0.7, 0)  -- (0.7, 2);
        \draw[very thick] (-0.46, -0.13) -- (-0.46, 1.87); \draw (0.46, -0.13)  -- (0.46, 1.87); \draw[dashed, gray!60] (0, 0.175) -- (0, 2.175);
    \end{scope}

    \begin{scope}[xshift=3cm, yshift=2cm]
        \draw (0, 2) ellipse (0.7cm and 0.175cm);
        \node[above] at (-0.05, 1){$t_4$}; \node[above] at (-0.7, 2.05){$1$}; \node[above] at (0.7, 2.05){$1$};
        \filldraw[black] (-0.7, 0) circle (1.5 pt); \filldraw[black] (0.7, 0) circle (1.5 pt); 
        \node[below, font=\scriptsize, inner sep=1pt] at (0.8, -.1) { $\mathcal{V}_{k_4}^{\alpha_4}$};
        \draw (-0.7, 0) arc (180:360:0.7cm and 0.175cm); \draw [dashed](0.7, 0) arc (0:180:0.7cm and 0.175cm);
        \draw (-0.7, 0) -- (-0.7, 2); \draw (0.7, 0)  -- (0.7, 2);
        \draw (-0.46, -0.13) -- (-0.46, 1.87); \draw[very thick] (0.46, -0.13)  -- (0.46, 1.87); \draw[dashed, gray!60] (0, 0.175) -- (0, 2.175);
    \end{scope}

    \begin{scope}[xshift=3cm, yshift=-2cm]
        \draw (0, 2) ellipse (0.7cm and 0.175cm);
        \node[above] at (-0.05, 1){$t_3$}; \node[above] at (-0.7, 2.05){$1$}; \node[above] at (0.7, 2.05){$1$};
        \filldraw[black] (-0.7, 0) circle (1.5 pt); \filldraw[black] (0.7, 0) circle (1.5 pt); 
        \node[below, font=\scriptsize, inner sep=1pt] at (0.8, -.1) { $\mathcal{V}_{k_3}^{\alpha_3}$};
        \draw (-0.7, 0) arc (180:360:0.7cm and 0.175cm); \draw [dashed](0.7, 0) arc (0:180:0.7cm and 0.175cm);
        \draw (-0.7, 0) -- (-0.7, 2); \draw (0.7, 0)  -- (0.7, 2);
        \draw (-0.46, -0.13) -- (-0.46, 1.87); \draw[very thick] (0.46, -0.13)  -- (0.46, 1.87); \draw[dashed, gray!60] (0, 0.175) -- (0, 2.175);
    \end{scope}

    \draw[->, line width=1pt] (4, 0.5) -- (5, 0.5);

    \begin{scope}[xshift=6.5cm, yshift=0cm]
        \draw (0, 1.5) ellipse (1.2cm and 0.4cm);
        \draw (-1.2, -1.5) arc (180:360:1.2cm and 0.4cm);
        \draw[dashed, gray] (-1.2, -1.5) arc (180:0:1.2cm and 0.4cm);
        
        \draw [very thick]  (-1.2, -1.5) -- (-1.2, 1.5) node[pos=0.6] {$$};
        \draw [very thick]  (1.2, -1.5) -- (1.2, 1.5)node[pos=0.6] {$$};
        
        \draw[very thick] (-0.3, -1.9) -- (-0.3, 1.1) node[pos=0.6] {$$};
        \draw[dashed, very thick] (-0.1, -1.1)  -- (-0.1, 1.9) node[pos=0.6] {$$};
        \draw[very thick] (0.3, -1.9) -- (0.3, 1.1)node[pos=0.6] {$$} ;
        
        \node[font=\footnotesize] at (-0.9, 2) {$t_1+1$};
        \filldraw[black] (-.9, -1.25) circle (1.5 pt);
        \node at  (-.85, -.9) {$\mathcal{V}_{k_1}^{\alpha_1}$};
        \node[font=\footnotesize] at (-0.75, 1.4) {$t_2+1$};
        \filldraw[black] (-.9, -1.75) circle (1.5 pt);
        \node at  (-.85, -2.2) {$\mathcal{V}_{k_2}^{\alpha_2}$};
        \node[font=\footnotesize] at (0.1, 1.3) {$t_5$};
        \node[font=\footnotesize] at (0.8, 1.4) {$t_3+1$};
         \node at  (.85, -2.2) {$\mathcal{V}_{k_3}^{\alpha_3}$};
        \filldraw[black] (.9, -1.75) circle (1.5 pt);
        \node[font=\footnotesize] at (0.85, 2) {$t_4+1$};
         \node at  (.85, -.9) {$\mathcal{V}_{k_4}^{\alpha_4}$};
        \filldraw[black] (.9, -1.25) circle (1.5 pt);
    \end{scope}

    \end{tikzpicture}%
}
\newcommand{\GinvC}{%
    \begin{tikzpicture}[baseline=(current bounding box.center), scale=2]

  \draw[dashed, gray] (-0.5, 0) arc (180:0:0.5 and 0.15);
  \draw (-0.5, 0) arc (180:360:0.5 and 0.15);

  \draw[->, >=stealth, thick] (-0.5, -0.1) arc (180:360:0.5 and 0.15) 
    node[midway, below, black] {$1$};

  \draw (-0.5, 0) -- (-0.5, 1.5);
  \draw (0.5, 0)  -- (0.5, 1.5);

  \draw[dashed, gray] (-0.5, 1.5) arc (180:0:0.5 and 0.15);
  \draw (-0.5, 1.5) arc (180:360:0.5 and 0.15);

  \draw (-0.5, 1.5) arc (180:0:0.5);

  \fill (.5,0) circle (1.2pt);   
  \node[right] at (.5,0) {\(\mathcal{V}_{k}^{\alpha}\)};
  \fill (0,2.0) circle (1.2pt); 
 \node[above] at (0,2) {\(V_k\)};
  \draw[<->, >=stealth, gray] (-0.8, 0) -- (-0.8, 1.5) node[midway, left, black] {$h$};

    \end{tikzpicture}%
}

\newcommand{\Gincorr}{%
    \begin{tikzpicture}[baseline=(current bounding box.center), scale=1]

    \draw[gray, line width=1pt] (-.9, -1.3) -- (-1.5,3);
    \draw[gray, line width=1pt] (-.9, -1.75) -- (-2.7, -1);
    \draw[gray, line width=1pt] (.9, -1.3) -- (2.5, 3);
    \draw[gray, line width=1pt] (.9, -1.75) -- (3.7,-.975);

    \begin{scope}[xshift=0cm, yshift=0cm]
       \draw (-1.2, 1.5) arc (180:360:1.2cm and 0.4cm);
        \draw[dashed, gray] (-1.2, 1.5) arc (180:0:1.2cm and 0.4cm);
        \draw (-1.2, -1.5) arc (180:360:1.2cm and 0.4cm);
        \draw[dashed, gray] (-1.2, -1.5) arc (180:0:1.2cm and 0.4cm);
        \draw (-1.2, 1.5) arc (180:0:1.2cm); 
        \draw [very thick]  (-1.2, -1.5) -- (-1.2, 1.5);
        \draw[<->, >=stealth, gray] (-1.5, -1.5) -- (-1.5, 1.5) node[midway, left, black] {$h$};
        \draw [very thick]  (1.2, -1.5) -- (1.2, 1.5);
        \draw[very thick] (-0.3, -1.9) -- (-0.3, 1.1);
        \draw[dashed, very thick] (-0.1, -1.1)  -- (-0.1, 1.9);
        \draw[very thick] (0.3, -1.9) -- (0.3, 1.1);
        
        \node[font=\footnotesize] at (-0.9, 2) {$t_1+1$};
        \filldraw[black] (-.9, -1.25) circle (1.5 pt);
        \node[font=\footnotesize] at (-0.75, 1.4) {$t_2+1$};
        \filldraw[black] (-.9, -1.75) circle (1.5 pt);
        \node[font=\footnotesize] at (0.1, 1.3) {$t_5$};
        \node[font=\footnotesize] at (0.8, 1.4) {$t_3+1$};
        \filldraw[black] (.9, -1.75) circle (1.5 pt);
        \node[font=\footnotesize] at (0.85, 2) {$t_4+1$};
        \filldraw[black] (.9, -1.25) circle (1.5 pt);

        \begin{scope}[xshift=-3.2cm, yshift=-1.0cm]
            \draw[dashed, gray] (-0.5, 0) arc (180:0:0.5 and 0.15);
            \draw (-0.5, 0) arc (180:360:0.5 and 0.15);
            \draw[->, >=stealth, thick] (-0.5, -0.1) arc (180:360:0.5 and 0.15) node[midway, below, black] {$1$};
            \draw (-0.5, 0) -- (-0.5, 1.5); \draw (0.5, 0)  -- (0.5, 1.5);
            \draw[dashed, gray] (-0.5, 1.5) arc (180:0:0.5 and 0.15);
            \draw (-0.5, 1.5) arc (180:360:0.5 and 0.15);
            \draw (-0.5, 1.5) arc (180:0:0.5);
            \fill (.5,0) circle (1.2pt);
            \fill (0,2.0) circle (1.2pt); \node[above] at (0,2) {\(V_{k_2}^2\)};
            \draw[<->, >=stealth, gray] (-0.8, 0) -- (-0.8, 1.5) node[midway, left, black] {$h$};
        \end{scope}

        \begin{scope}[xshift=3.2cm, yshift=-1.0cm]
            \draw[dashed, gray] (-0.5, 0) arc (180:0:0.5 and 0.15);
            \draw (-0.5, 0) arc (180:360:0.5 and 0.15);
            \draw[->, >=stealth, thick] (-0.5, -0.1) arc (180:360:0.5 and 0.15) node[midway, below, black] {$1$};
            \draw (-0.5, 0) -- (-0.5, 1.5); \draw (0.5, 0)  -- (0.5, 1.5);
            \draw[dashed, gray] (-0.5, 1.5) arc (180:0:0.5 and 0.15);
            \draw (-0.5, 1.5) arc (180:360:0.5 and 0.15);
            \draw (-0.5, 1.5) arc (180:0:0.5);
            \fill (.5,0) circle (1.2pt);
            \fill (0,2.0) circle (1.2pt); \node[above] at (0,2) {\(V_{k_3}^3\)};
            \draw[<->, >=stealth, gray] (-0.8, 0) -- (-0.8, 1.5) node[midway, left, black] {$h$};
        \end{scope}

        \begin{scope}[xshift=-2.0cm, yshift=3.0cm]
            \draw[dashed, gray] (-0.5, 0) arc (180:0:0.5 and 0.15);
            \draw (-0.5, 0) arc (180:360:0.5 and 0.15);
            \draw[->, >=stealth, thick] (-0.5, -0.1) arc (180:360:0.5 and 0.15) node[midway, below, black] {$1$};
            \draw (-0.5, 0) -- (-0.5, 1.5); \draw (0.5, 0)  -- (0.5, 1.5);
            \draw[dashed, gray] (-0.5, 1.5) arc (180:0:0.5 and 0.15);
            \draw (-0.5, 1.5) arc (180:360:0.5 and 0.15);
            \draw (-0.5, 1.5) arc (180:0:0.5);
            \fill (.5,0) circle (1.2pt);
            \fill (0,2.0) circle (1.2pt); \node[above] at (0,2) {\(V_{k_1}^1\)};
            \draw[<->, >=stealth, gray] (-0.8, 0) -- (-0.8, 1.5) node[midway, left, black] {$h$};
        \end{scope}

        \begin{scope}[xshift=2.0cm, yshift=3.0cm]
            \draw[dashed, gray] (-0.5, 0) arc (180:0:0.5 and 0.15);
            \draw (-0.5, 0) arc (180:360:0.5 and 0.15);
            \draw[->, >=stealth, thick] (-0.5, -0.1) arc (180:360:0.5 and 0.15) node[midway, below, black] {$1$};
            \draw (-0.5, 0) -- (-0.5, 1.5); \draw (0.5, 0)  -- (0.5, 1.5);
            \draw[dashed, gray] (-0.5, 1.5) arc (180:0:0.5 and 0.15);
            \draw (-0.5, 1.5) arc (180:360:0.5 and 0.15);
            \draw (-0.5, 1.5) arc (180:0:0.5);
            \fill (.5,0) circle (1.2pt);
            \fill (0,2.0) circle (1.2pt); \node[above] at (0,2) {\(V_{k_4}^4\)};
           
            \draw[<->, >=stealth, gray] (-0.8, 0) -- (-0.8, 1.5) node[midway, left, black] {$h$};
        \end{scope}
    \end{scope}

    \draw[->, line width=1.5pt, >=stealth] (4.8, 0) -- (6.0, 0);

    \begin{scope}[xshift=8.2cm, yshift=0cm]
        \draw (-1.2, 1.5) arc (180:360:1.2cm and 0.4cm);
        \draw[dashed, gray] (-1.2, 1.5) arc (180:0:1.2cm and 0.4cm);
        \draw (-1.2, -1.5) arc (180:360:1.2cm and 0.4cm);
        \draw[dashed, gray] (-1.2, -1.5) arc (180:0:1.2cm and 0.4cm);
        \draw (-1.2, 1.5) arc (180:0:1.2cm); 
        \draw [very thick]  (-1.2, -1.5) -- (-1.2, 1.5);
        \draw [very thick]  (1.2, -1.5) -- (1.2, 1.5);
        \draw[very thick] (-0.3, -1.9) -- (-0.3, 1.1);
        \draw[dashed, very thick] (-0.1, -1.1)  -- (-0.1, 1.9);
        \draw[very thick] (0.3, -1.9) -- (0.3, 1.1);
          \draw[<->, >=stealth, gray] (-1.5, -1.5) -- (-1.5, 1.5) node[midway, left, black] {$h$};
        \node[font=\footnotesize] at (-0.9, 1.5) {$t_1$};
        \filldraw[black] (-.65, 2.3) circle (1.5 pt);
        \node[above] at  (-1, 2.3) {${V}_{k_1}^{1}$};
        \node[font=\footnotesize] at (-0.75, .9) {$t_2$};
        \filldraw[black] (-.9, 1.9) circle (1.5 pt);
        \node at  (-1.4, 2) {${V}_{k_2}^{2}$};
        \node[font=\footnotesize] at (0.1, .8) {$t_5$};
        \node[font=\footnotesize] at (0.8, .9) {$t_3$};
        \node at  (1.4, 1.9) {${V}_{k_3}^{3}$};
        \filldraw[black] (.9, 1.9) circle (1.5 pt);
        \node[font=\footnotesize] at (0.85, 1.5) {$t_4$};
        \node[above] at  (1.05, 2.3) {${V}_{k_4}^{4}$};
        \filldraw[black] (.7, 2.3) circle (1.5 pt);
    \end{scope}

\end{tikzpicture}
}
\newcommand{\Cylplumb}{%
\begin{tikzpicture}[baseline=(current bounding box.center), scale=1.1]

    \begin{scope}[xshift=0cm, yshift=0.5cm]
        \draw[thick] (0,1.2) ellipse (0.6cm and 0.18cm);
        \draw[thick] (0,-1.2) ellipse (0.6cm and 0.18cm);
        
        \draw[thick] (-0.6,-1.2) -- (-0.6,1.2);
        \draw[thick] (0.6,-1.2) -- (0.6,1.2);
        
        \draw[<->, >=stealth] (0.9,-1.2) -- (0.9,1.2) node[midway, right, black] {$h$};
        
        \node[below] at (0, -1.5) { $\tau$};
    \end{scope}

    \draw[->, line width=1pt, >=stealth] (1.6, 0.5) -- (3, 0.5) 
        node[midway, above, black] {$w_1 = e^{\frac{2\pi i z}{\tau}}$};

    \begin{scope}[xshift=5cm, yshift=0.5cm]
        \draw[thick, even odd rule] (0,0) circle (1.4cm) (0,0) circle (0.45cm);
        \draw[thick] (0,0) circle (1.4cm);
        \draw[thick] (0,0) circle (0.45cm);
        
        \node[ black] at (0,-.6) {$|w_1|=e^{-\frac{2\pi h}{\tau}}$};
        \node[  black] at (0,1.6){$|w_1|=1$};
        
    \end{scope}

    \draw[<-, line width=1pt, >=stealth, ] (7, 0.5) -- (9.5, 0.5)
        node[midway, below, black, font=\footnotesize] {plumbing};

    \begin{scope}[xshift=10.8cm, yshift=0.5cm]
        \draw[thick] (0,-2) ellipse (1.4cm and 1.4cm);
         \fill (0,-2) circle (1.2pt);
        \draw[thick] (0,2) ellipse (1.4cm and 1.4cm);
         \fill (0,2) circle (1.2pt);
        
        \draw[dashed,thick] (0,2) ellipse (0.7cm and 0.7cm);
         \draw[dashed,thick] (0,-2) ellipse (0.7cm and 0.7cm);
        
        \node[right] at (1.1, 3.1) {$|w_1| = 1$};
        \node[right] at (1.1, -1) {$|w_2| = 1$};
        
        \node[above] at (-2.5, 0) {$w_1 w_2 = e^{-\frac{2\pi h}{\tau}}$};
        
         \node[ black] at (0.05,1.2) {$|w_1|=e^{-\frac{\pi h}{\tau}}$};
          \node[ black] at (0.1,-2.85) {$|w_2|=e^{-\frac{\pi h}{\tau}}$};
    \end{scope}

\end{tikzpicture}

}

\newcommand{\drawRiemannSurface}{%
  \begin{tikzpicture}[scale=1.1, line join=round, line cap=round]
    \draw[thick] (-2.2, 0.8) -- (-2.2, -3.5);
    \draw[thick] (-1.4, 0.8) -- (-1.4, -3.5);
    \draw[thick] (-2.2, -3.5) to[out=-20,in=-160] (-1.4, -3.5);
    \draw[thick, dashed] (-2.2, -3.5) to[out=20,in=160] (-1.4, -3.5);
    \node at (0, -1.8) {$\dots$};
    \node[below] at (0, -2.1) {b cylinders};
    \draw[thick] (1.4, 0.8) -- (1.4, -3.5);
    \draw[thick] (2.4, 0.8) -- (2.4, -3.5);
    \draw[thick] (1.4, -3.5) to[out=-20,in=-160] (2.4, -3.5);
    \draw[thick, dashed] (1.4, -3.5) to[out=20,in=160] (2.4, -3.5);
    \draw[thick] (-3.2, -0.6) to[out=110,in=-90] (-3.6, 0.2) to[out=90,in=180] (-2.9, 0.9) to[out=0,in=180] (-2.2, 0.8);
    \draw[thick] (-1.4, 0.8) to[out=0,in=180] (0, 0.6) to[out=0,in=180] (1.4, 0.8);
    \draw[thick] (2.4, 0.8) to[out=0,in=90] (3.6, 0.2) to[out=-90,in=70] (3.2, -0.6);
    \draw[thick] (-2.2, 0.8) to[out=-20,in=-160] (-1.4, 0.8);
    \draw[thick, dashed] (-2.2, 0.8) to[out=20,in=160] (-1.4, 0.8);
    \draw[thick] (1.4, 0.8) to[out=-20,in=-160] (2.4, 0.8);
    \draw[thick, dashed] (1.4, 0.8) to[out=20,in=160] (2.4, 0.8);
    \draw[thick] (-0.9, 0.1) to[out=-30,in=-150] (-0.3, 0.1);
    \draw[thick] (-0.8, 0.04) to[out=30,in=150] (-0.4, 0.04);
    \node[above] at (-0.6, 0.1) {1};
    \node at (0, 0.08) {$\dots$};
    \draw[thick] (0.3, 0.1) to[out=-30,in=-150] (0.9, 0.1);
    \draw[thick] (0.4, 0.04) to[out=30,in=150] (0.8, 0.04);
    \node[above] at (0.6, 0.1) {g};
    \node at (-3.1, 0.2) {$\times$}; 
    \node[above] at (-3.1, 0.3) {1};
    \node at (3.1, 0.2) {$\times$};  
    \node[above] at (3.1, 0.3) {n};
    \draw[thick] (-3.2, -0.6) to[out=-10,in=-170] (3.2, -0.6);
    \draw[thick, dashed] (-3.2, -0.6) to[out=10,in=170] (3.2, -0.6);
    \draw[thick] (-3.2, -0.6) -- (-3.2, -3.5);
    \draw[thick] (3.2, -0.6)  -- (3.2, -3.5);
    \draw[thick] (-3.2, -3.5) to[out=-10,in=-170] (3.2, -3.5);
    \draw[thick, dashed] (-3.2, -3.5) to[out=10,in=170] (3.2, -3.5);
  \end{tikzpicture}%
}

\title{\boldmath  Quantum Gravity in Asymptotically Flat Spacetimes from Unstable D-Branes}

\author[a,b]{Roji Pius}

\affiliation[a]{The Institute of Mathematical Sciences,\\IV Cross Road, C.I.T. Campus, Taramani,
Chennai, India-600113}
\affiliation[b]{Homi Bhabha National Institute,\\Training School Complex, Anushakti Nagar, Mumbai,
India 400094}
\emailAdd{rojipius@imsc.res.in}

\abstract{We present a non-perturbative formulation of quantum gravity in asymptotically flat spacetimes using Witten's cubic open string field theory (OSFT) on unstable D-branes. Given that the massless spin-2 sector of the closed string spectrum inherently encodes the degrees of freedom of quantum gravity, we establish this connection by mapping on-shell closed string excitations directly to gauge-invariant operators in OSFT, realizing a concrete flat-space analogue of the AdS/CFT correspondence. By evaluating the OSFT correlation functions of these operators perturbatively around the tachyon vacuum, we systematically reproduce the all-orders genus expansion of the closed string \(S\)-matrix for the associated states. This demonstrates that closed string theory-and thus quantum gravity-emerges dynamically from the open string tachyon vacuum, establishing Witten's OSFT as a fully calculable, non-perturbative framework for flat-space quantum gravity.  }

\begin{document} 
\maketitle
\flushbottom

\section{Introduction}
\label{sec:intro}

A calculable, non-perturbative formulation of quantum gravity in asymptotically flat spacetimes remains one of the most ambitious goals in theoretical physics. Although flat-space quantum gravity can, in principle, be defined directly through Closed String Field Theory (CSFT), its practical realization has been notoriously difficult due to the highly intricate, non-polynomial structure of the action and the mathematically punishing nature of non-local closed string vertices \cite{Zwiebach:1992ie, Sen:2015uaa, Moosavian:2017qsp, Moosavian:2017sev, Costello:2019fuh}. Conversely, while the AdS/CFT correspondence provides a mathematically rigorous, non-perturbative framework for quantum gravity in Anti-de Sitter space \cite{Maldacena:1997re, Witten:1998qj, Gubser:1998bc}, a compelling and universally accepted holographic dictionary for asymptotically flat spacetimes has remained elusive. \par

In this paper, we exploit an alternative approach rooted in the dual nature of string theory: open string loop amplitudes fundamentally encode closed string states as internal poles. This open-closed duality hints at a profound possibility: closed string degrees of freedom may be formally identified with gauge-invariant operators in the open string theory, while the true closed string vacuum emerges as a particular classical solution to the open string field equations. If this realization holds, the full, non-perturbative closed string S-matrix can be directly constructed from the correlation functions of these gauge-invariant operators evaluated around the appropriate open string background. \par

Unlike its closed counterpart, open string theory possesses an elegant, highly calculable field-theoretic formulation governed by Witten’s cubic classical action \cite{Witten:1985cc}. The Feynman rules derived from this single cubic vertex successfully reproduce the Veneziano amplitude at tree level \cite{Giddings:1986iy} and, more fundamentally, provide a precise geometric triangulation of the moduli space of Riemann surfaces with arbitrary numbers of genera, boundaries, and boundary punctures \cite{Giddings:1986wp}. Crucially, the most profound application of Open String Field Theory (OSFT) to date has been the verification of Sen’s celebrated tachyon condensation conjectures \cite{Sen:1998ii, Sen:1998sm, Sen:1999nx, Berkovits:2000hf}. \par

The presence of a tachyon in the open string spectrum is a consequence of defining the theory at the unstable maximum of the tachyon potential. Upon condensing to the local minimum—the tachyon vacuum—the tachyon completely disappears from the physical spectrum. A cornerstone of Sen’s conjectures is that this non-perturbative tachyon vacuum represents the true closed string vacuum, devoid of any physical open string degrees of freedom. This physical picture strongly implies that if closed string excitations can be successfully mapped to gauge-invariant operators in OSFT, their correlation functions evaluated at the tachyon vacuum should precisely reconstruct closed string scattering amplitudes. \par

Remarkably, it is possible to construct gauge-invariant open string field operators, denoted as \(\mathcal{O}_V(\Psi)\), which are parameterized by bulk worldsheet vertex operators V of conformal dimension (1,1) \cite{Hashimoto:2001sm}. These operators originate from an open-to-closed transition vertex that naturally arises when analyzing closed string factorization in OSFT loop diagrams \cite{Shapiro:1987gq, Shapiro:1987ac, Freedman:1987fr}. A systematic geometric study of this transition vertex demonstrated that the moduli spaces of surfaces involving both open and closed string punctures can be fully covered by combining this transition vertex with the standard cubic open string vertex \cite{Zwiebach:1992bw}. Around the perturbative vacuum, the perturbative expansion of the correlation functions of these gauge-invariant operators yields the standard S-matrix elements of the corresponding closed string vertex operators, integrated over the moduli spaces of surfaces with boundaries. Consequently, the operators \(\mathcal{O}_{V}\) can be unambiguously identified with the closed string states represented by V. \par

In this work, we demonstrate that evaluating the correlation functions of these gauge-invariant operators in OSFT defined on a space-filling D25-brane around the tachyon vacuum yields the perturbative, flat-space closed string S-matrix, expressed as an integration over the moduli spaces of Riemann surfaces without boundaries. This involves constructing the perturbation theory around the tachyon vacuum using analytic background solutions \cite{Schnabl:2005gv, Okawa:2006vm, Erler:2009uj} and tracking the underlying worldsheet geometries at each order in the perturbative expansion. \par
 
The worldsheets generated by this perturbation theory emerge with closed string punctures and infinitely long boundary cylinders subject to D25-brane boundary conditions. Following the general prescription for boundary closure introduced in \cite{Gaiotto:2001ji}, we interpret these worldsheets as geometric configurations constructed via the plumbing-fixture gluing of single, closed-string-punctured disks to a worldsheet containing exclusively bulk punctures. Then, by invoking the soft theorems of the zero-momentum dilaton and graviton \cite{Bergman:1994qq, Sen:2017xjn} and performing the systematic summation over boundaries, we demonstrate that the OSFT correlation function evaluated around the tachyon vacuum reduces precisely to the all-orders genus expansion of the perturbative closed string S-matrix. \par

The remainder of this paper is organized as follows. In Section \ref{sec:OSFT}, we provide a brief overview of Witten's formulation of open string field theory. Section \ref{sec:GInv} is dedicated to the construction of gauge-invariant operators in OSFT and their mapping onto the non-linear, gauge-invariant spacetime structures of the low-energy effective theory. In Section \ref{sec:Condensation}, we review the mechanism through which the closed string vacuum emerges as a classical solution of OSFT. In Section \ref{sec:Perturbation}, we develop the formal perturbation theory around the non-perturbative tachyon vacuum background. In Section \ref{sec:Smatrix}, we present the core proof of this work, demonstrating that the correlation functions of gauge-invariant operators evaluated perturbatively around the tachyon vacuum precisely recover the all-orders genus expansion of the pure closed string S-matrix. Finally, in Section \ref{sec:Discussion}, we conclude with a brief discussion of our findings and outline several compelling directions for future research.

\section{Witten's cubic open string field theory}\label{sec:OSFT}

Open string field theory (OSFT), governed by a gauge-invariant cubic action introduced by Witten, provides a non-perturbative formulation of open bosonic string theory \cite{Witten:1985cc}. Unlike first-quantized string theory, which directly defines the perturbative expansion of \(S\)-matrix elements via a worldsheet conformal field theory (CFT), OSFT defines an off-shell action for a string field \(\vert{}\Psi\rangle\). This string field represents a state in the first-quantized open string Hilbert space and contains an infinite tower of classical particle fields. \par

The underlying worldsheet CFT acts as the kinematic foundation of OSFT, defining the off-shell string field Hilbert space and dictating the algebraic operations used to construct the action. Because OSFT is a background-dependent formulation, it is built upon a specific, fixed choice of a reference boundary conformal field theory (BCFT). This BCFT consists of two decoupled sectors: the matter sector and the ghost sector. In the standard flat-space background, the matter sector is a BCFT of 26 free scalar fields \(X^\mu(\sigma, \tau)\) for \(\mu=0,\cdots, 25\). For a space-filling \(D25\)-brane, these fields satisfy Neumann boundary conditions (\(\partial_\sigma X^\mu = 0\)) across all 26 coordinates. The ghost sector is the universal worldsheet reparametrization ghost system, consisting of the anticommunicating \(b\) field of conformal weight 2 and the \(c\) field of conformal weight \(-1\). \par

\subsection{String field}
\begin{figure}[t!]
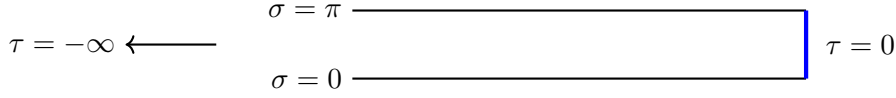

    \centering
    \spacetimestrip
    \caption{The world-sheet of a free string is an infinitely long flat strip with spatial coordinates $\sigma \in [0, \pi]$ and time coordinate $\tau \in (-\infty, 0]$.} \label{Fstring}
\end{figure}

The string field \(\vert{}\Psi\rangle\) is an element of this total BCFT Hilbert space restricted to ghost number 1. The ghost number is a conserved charge arising from a \(U(1)\) symmetry in the reparametrization ghost sector, and it acts as the primary selection rule for physical states, gauge parameters, and valid CFT correlation functions. The ghost number operator is the conserved charge obtained by integrating the ghost current along a closed contour around the origin:
\begin{equation}\label{gno} N_{g}=\oint \frac{dz}{2\pi i} b(z)c(z). \end{equation}
By evaluating the Operator Product Expansions (OPEs) of the ghost current with the fundamental fields, we find that the \(c(z)\) field has ghost number \(+1\), the \(b(z)\) field has ghost number \(-1\), and the vacuum state \(\vert{}0\rangle\) has ghost number \(0\). \par
Under the state-operator correspondence, the string field corresponds to a local boundary vertex operator \(\mathcal{V}(0)\) inserted at the origin:
\begin{equation} 
\vert{}\Psi \rangle =\mathcal{V}(0)\vert{}0\rangle. 
\end{equation} 
Expanding this state in the BCFT mode basis yields an infinite tower of spacetime component fields: 
\begin{equation} 
\vert{}\Psi \rangle =\int d^{26}p\left[T(p)c_{1}+A_{\mu }(p)\alpha _{-1}^{\mu }c_{1}-i\sqrt{\frac{1}{2}}B(p)c_{0}+\dots \right]\vert{}p\rangle, 
\end{equation}
where \(T(p)\) represents the open-string spacetime tachyon field and \(A_\mu(p)\) represents the photon gauge field. Because both \(c_1\vert{}0\rangle\) and \(c_0\vert{}0\rangle\) possess a ghost number of \(+1\), the spacetime tachyon \(T(p)\), the gauge field \(A_\mu(p)\), and the Lautrup–Nakanishi auxiliary field \(B(p)\) all reside in the ghost-number-1 sector. \par
\subsection{Integration, star product and the cubic action}

\begin{figure}[t!]
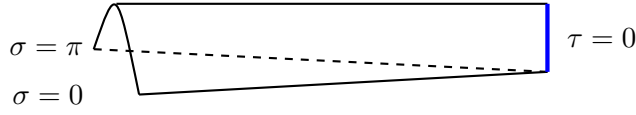

    \centering
    \integration
    \caption{The the integration functional corresponds to folding a single open string strip in half and gluing its left half (\(0 \le \sigma \le \frac{\pi}{2}\)) to its right half (\(\frac{\pi}{2} \le \sigma \le \pi\)). }\label{integration}
\end{figure}
The two non-local operations required to define Witten's action—the star product (\(\ast \)) and the integration functional (\(\int \))—are uniquely determined by specific conformal maps on the worldsheet. The worldsheet of a non-interacting free string is an infinitely long flat strip with spatial coordinate \(\sigma \in [0, \pi]\) and time coordinate \(\tau \in (-\infty, 0]\), as illustrated in Figure~\ref{Fstring}. Geometrically, the integration functional corresponds to folding a single open string strip in half and gluing its left half (\(0 \le \sigma \le \frac{\pi}{2}\)) to its right half (\(\frac{\pi}{2} \le \sigma \le \pi\)); see Figure~\ref{integration}. The star product \(\Psi_1 \ast \Psi_2\) glues the right half of the first string to the left half of the second string. This operation satisfies the cyclicity property:
\begin{equation}\label{cyclicity}
 \int \Psi_1 \ast \Psi_2 = (-1)^{\text{gh}(\Psi_1)\text{gh}(\Psi_2)} \int \Psi_2 \ast \Psi_1, 
 \end{equation}
where \(\text{gh}(\Psi)\) denotes the ghost number of the string field \(\Psi \). \par
\begin{figure}[t!]
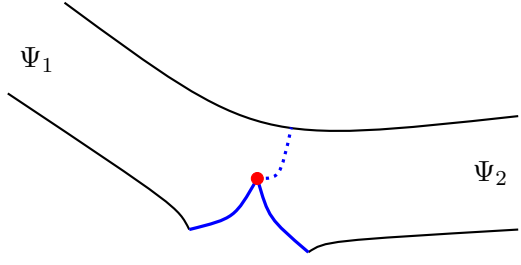

    \centering
    \starproduct
    \caption{The star product \(\Psi_1 \ast \Psi_2\) glues the right half of string 1 to the left half of string 2. }\label{starproduct}
\end{figure}

\begin{figure}[t!]
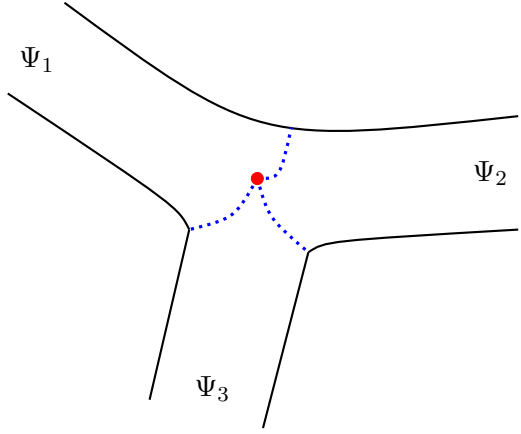

    \centering
    \wittenvertex
    \caption{The three string interaction is constructed by identifying and gluing the right half of String 1 to the left half of String 2, the right half of String 2 to the left half of String 3, and the right half of String 3 to the left half of String 1. The exact midpoint \(\sigma=\frac{\pi}{2}\) is shared identically by all three strings. This creates a conical singularity at the midpoint  with an excess angle of \(\pi\). } \label{wittenvertex}
\end{figure}

The worldsheet BRST charge, which is derived from the matter and ghost sector stress tensors of the BCFT, is defined as:
\begin{equation}\label{BRSTOp} 
Q_{B}=\oint \frac{dz}{2\pi i}c(z)\left(T_{\text{matter}}(z)+\frac{1}{2}T_{\text{ghost}}(z)\right). 
\end{equation}
This operator satisfies the Leibniz rule with respect to the star product:
\begin{equation}\label{LruleQB}
 Q_B(\Psi_1 \ast \Psi_2) = (Q_B\Psi_1) \ast \Psi_2 + (-1)^{\text{gh}(\Psi_1)} \Psi_1 \ast Q_B\Psi_2. 
 \end{equation}
Using these definitions of integration and the star product, the classical action for OSFT can be expressed as:
\begin{equation}\label{OSFTAction} S[\Psi] = \frac{1}{2} \int \Psi \ast Q_B \Psi + \frac{g_o}{3} \int \Psi \ast \Psi \ast \Psi, \end{equation}
where \(g_{o}\) is the open string coupling constant. The cubic interaction term in the action is constructed by gluing the right half of the first string to the left half of the second string, the right half of the second string to the left half of the third string, and the right half of the third string to the left half of the first string. The exact midpoint \(\sigma=\frac{\pi}{2}\) is shared identically by all three strings, creating a conical singularity with an excess angle of \(\pi \) at the junction where the three worldsheets meet. \par
The nilpotency of the BRST operator (\(Q_B^2 = 0\)) relies strictly on the cancellation of the total central charge between the matter and ghost sectors. This nilpotency ensures the invariance of the classical action under the infinite-dimensional gauge transformation:
\begin{equation}\label{gauget} 
\delta \Psi =Q_{B}\Lambda +g_o(\Psi \ast \Lambda -\Lambda \ast \Psi ), 
\end{equation}
where \(\Lambda \) is a gauge parameter state possessing ghost number 0. \par

\section{Gauge invariant OSFT operators around the perturbative vacuum}\label{sec:GInv}

Perturbation theory in Witten’s OSFT provides a systematic framework for calculating string scattering amplitudes. By implementing the Siegel gauge,
\begin{equation}\label{sgauge} 
b_0\Psi = 0, 
\end{equation}
the infinite-dimensional gauge invariance is broken, yielding a well-defined set of Feynman rules. The propagator handles the free evolution of an open string strip, inverting the kinetic operator under the Siegel constraint:
\begin{equation}\label{SGprop} 
P=\frac{b_{0}}{L_{0}}=b_{0}\int _{0}^{\infty }dt\,e^{-tL_{0}}. 
\end{equation}
The Schwinger parameter \(t\) represents the length of the string worldsheet strip, and the antighost zero mode \(b_{0}\) acts as a differential form integrator over the moduli space of the strip. The cubic vertex glues three string strips together via Witten's star product; see Figure~\ref{wittenvertex}. \par
The worldsheet constructed by the cubic vertex comes equipped with a metric induced by the following Strebel differential on the upper half-plane (UHP) with three boundary punctures:
\begin{equation}\label{strebeld3} 
\left(dw\right)^2= \frac{z^2-z+1}{z^2(1-z)^2}\left(dz\right)^2,
 \end{equation}
where \(w\) is the coordinate on the worldsheet and \(z\) is the coordinate on the UHP. In fact, an arbitrary worldsheet constructed from the Feynman rules of Witten's OSFT is equipped with a metric induced by a Strebel differential \cite{Giddings:1986iy, Giddings:1986wp}. For instance, a tree-level OSFT diagram with \(n\) external open strings carries the Strebel differential:
\begin{equation}\label{strebeldn}
 \left(dw\right)^2= \frac{\prod_{i=1}^{n-2}\left(z-Z_i\right)\left(z-\bar Z_i\right)}{\prod_{j=1}^n\left(z-X_j\right)^2}\left(dz\right)^2, 
 \end{equation}
where \(Z_{i}\) is the midpoint of the \(i^{\text{th}}\) cubic vertex and \(X_{j}\) corresponds to the asymptotic infinity of the \(j^{\text{th}}\) external open string. These locations can be determined in terms of the \((n-2)\) Schwinger parameters of the OSFT diagram, utilizing the fact that they correspond to the distances between the midpoints of cubic string vertices connected via OSFT propagators, measured with respect to the metric induced by the Strebel differential on the worldsheet. \par

\subsection{Open-to-closed string vertices }\label{GIO}

\begin{figure}[t!]
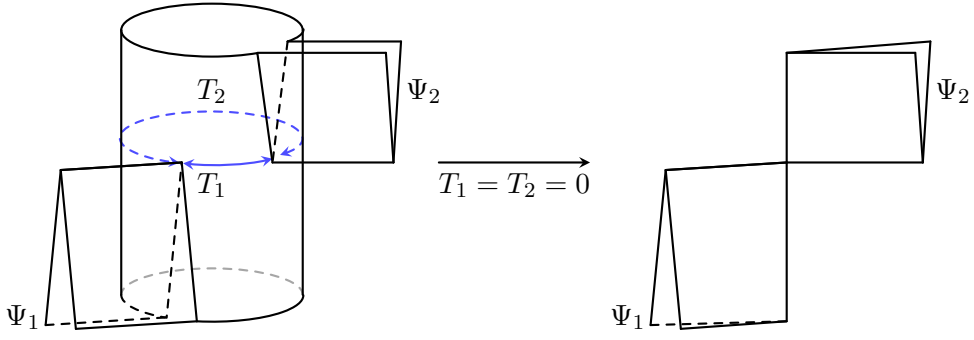
\label{tpointloop}
    \centering
    \tpointloop
    \caption{In the limit where the Schwinger parameters $T_1$ and $T_2$ vanish, the one loop OSFT diagram with two external open strings  can be interpreted as a worldsheet obtained by gluing a pair of open-to-closed string vertices using a closed string propagator. }
\end{figure}

Consider a one-loop diagram with two external open strings. The standard Feynman rules yield an annulus worldsheet with boundary insertions for the external states. In the limit of vanishing Schwinger parameters for the OSFT propagators, this diagram factorizes into a pair of open-closed string vertices connected by a closed string propagator. These vertices naturally define gauge-invariant OSFT operators parameterized by on-shell closed string vertex operators \(V\). The correlation functions of these gauge-invariant OSFT operators yield the scattering amplitudes for closed strings interacting with a \(D\)-brane background \cite{Koyama:2020qfb, Garousi:2004pi, Takahashi:2003kq}.

\begin{figure}[t!]
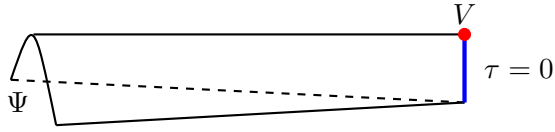

    \centering
    \opentclosed
    \caption{The  gauge invariant operator \(\mathcal{O}_V\) is obtained by inserting an on-shell bulk vertex operator \(V\) in the BCFT at the red dot. }
    \label{opentclosed}
\end{figure}
Mathematically, the OSFT gauge-invariant operator takes the following form:
\begin{equation}\label{GInvOperator} 
\mathcal{O}_{V}(\Psi )=\int V\left(\frac{\pi}{2}\right) \Psi, 
\end{equation}
where \(\Psi \) is the dynamical open string field and the on-shell bulk operator \(V\) is inserted precisely at the local midpoint coordinate \(\sigma = \frac{\pi}{2}\)—corresponding to the red dot in Figure~\ref{opentclosed}~\cite{Hashimoto:2001sm}. Under an open string gauge transformation~\eqref{gauget}, for the operator to be gauge-invariant (\(\delta \mathcal{O}_V = 0\)), the cancellations of the linear term (\(Q_B \Lambda\)) and the non-linear term (\(\Psi \ast \Lambda - \Lambda \ast \Psi\)) must occur separately. By pulling the BRST charge \(Q_{B}\) off the field via integration by parts, it acts directly on the bulk operator \(V\). Because the bulk operator is strictly chosen to be on-shell, it is annihilated by the BRST charge (\(Q_B V = 0\)). The non-linear, commutator-like term cancels identically due to the cyclic trace property and the fact that the bulk operator is inserted precisely at the midpoint of the open string at \(\tau=0\). \par
 
 \subsection{Non-linear gauge invariant spacetime operators from OSFT operators}
 
Let us demonstrate how linear stringy observables dynamically generate non-linear, gauge-invariant field theory operators. The low-energy behavior of the gauge-invariant operator~\eqref{GInvOperator} can be computed by incorporating it as a sourced component inside Witten’s action:
\begin{equation}\label{WactionS} 
S[\Psi ,J]=\frac{1}{2}\int \Psi \ast Q_{B}\Psi +\frac{g_o}{3}\int \Psi \ast \Psi \ast \Psi +J \cdot \int V \Psi, 
\end{equation}
where \(J\) acts as an external classical source coupling directly to the midpoint observable \cite{Zwiebach:1992bw, Koyama:2020qfb}. We choose the Siegel gauge~\eqref{sgauge} and split the full string field into its massless gauge boson component and an infinite tower of massive fluctuations \(\vert{}\chi\rangle\):
\begin{equation}\label{splitsfield} \Psi = A_\mu(x) \alpha_{-1}^\mu c_1 \vert{}0\rangle+\vert{}\chi\rangle. \end{equation}
Substituting this decomposition into the sourced action~\eqref{WactionS} isolates the kinetic and interaction terms involving the massive fields \(\vert{}\chi\rangle\):
\begin{equation}\label{chiappwaction} S[A,\chi ,J]=S_{\text{YM}}^{(3)}[A]+\frac{1}{2}\int \chi \ast Q_{B}\chi +g_o\int \chi \ast A\ast A+g_o\int \chi \ast \chi \ast A+\frac{g_o}{3}\int \chi \ast \chi \ast \chi +J\int V \chi, \end{equation}
where \(S_{\text{YM}}^{(3)}[A]\) represents the quadratic and cubic portions of the standard Yang-Mills action, which originate purely from the massless sector under the state-operator mapping. \par
To integrate out the massive field sector, we solve the classical equation of motion for \(\vert{}\chi\rangle\) under the Siegel gauge constraint:
\begin{equation}\label{chieom} \frac{\delta S}{\delta \chi }=0 \implies Q_{B}\chi +g_o A\ast A+2g_o A\ast \chi +g_o \chi \ast \chi  +JV=0. \end{equation}
We can solve this non-linear equation perturbatively in powers of the open string coupling \(g_{o}\) and the source \(J\). This is accomplished by acting on the equation of motion with the Siegel gauge propagator projected onto the massive sector:
\begin{equation}\label{P} \hat P = \frac{b_0}{L_0}\left(\mathbb{I}-\mathbb{P}\right), \end{equation}
where \(\mathbb{P}\) is the projection operator that projects a string field onto the massless sector of the worldsheet Hilbert space. Utilizing the property \(\{Q_B, b_0\} = L_0\), we obtain the recursive solution for the massive fields:
\begin{equation}\label{chisolved} \chi = - J V \hat P -g_o \hat P (A \ast A) + \mathcal{O}(g_o^2, J^2). \end{equation}
Plugging this explicit perturbative solution for \(\vert{}\chi\rangle\) back into the sourced action~\eqref{chiappwaction} and collecting all terms linear in the external source \(J\), we extract the low-energy effective operator \(\mathcal{O}_{\text{eff}}(V) = \left. \frac{\delta S_{\text{eff}}}{\delta J} \right\vert{}_{J=0}\):
\begin{equation}\label{effoperator} \mathcal{O}_{\text{eff}}(V)=\int V A-g_o\int V \left(\hat P (A\ast A)\right)+ \mathcal{O}(g_o^2). \end{equation}
Crucially, integrating out the massive fields injects higher-order non-linear contact interactions into the operator definition. \par
Let us explicitly compute the effective operator \(\mathcal{O}_{\text{eff}}(V_G)\) associated with an on-shell bulk graviton state with polarization \(h_{\mu \nu }\). The closed string vertex operator \(V_{G}\) for a graviton carrying polarization tensor \(h_{\mu\nu}(q)\) and momentum \(q^{\rho }\), inserted at the open string midpoint—which conformally maps to the point \(z = i\) in the upper half-plane (UHP)—is given by:
\begin{equation}\label{gravitonVOp} V_G(z,\bar z)=h_{\mu \nu }(q)c\bar c\partial X^{\mu }\bar \partial  X^{\nu }e^{iq\cdot X}\left(z,\bar z\right). \end{equation}
This bulk operator satisfies the standard physical state conditions dictated by the closed string BRST cohomology:
\begin{equation}\label{BRSTingraviton} q^{2}=0 \quad (\text{massless}), \quad q^{\mu }h_{\mu \nu }=0 \quad (\text{transverse}), \quad h^{\mu }{}_{\mu }=0 \quad (\text{traceless}). \end{equation}
Similarly, the boundary vertex operator representing the massless gauge boson is given by:
\begin{equation}\label{gaugeVo} V_A(z)=A_{\rho }(p)c\partial X^{\rho }e^{ip\cdot X}\left(z\right). \end{equation}
We first evaluate the linear overlap of a single gauge boson with the bulk graviton, which corresponds to the first term in~\eqref{effoperator}. This requires calculating a two-point mixed bulk-boundary CFT correlation function \(\langle V_G(i,-i) A(0) \rangle_{\text{UHP}}\). Due to spacetime translation invariance along the boundary directions, momentum must be strictly conserved: \(\delta^{26}(q + p) = 0\). Together with the on-shell condition~\eqref{BRSTingraviton} (\(q^2 = 0\)) and the transversality requirement, the kinematics force the polarization contraction to vanish identically:
\begin{equation}\label{Lvanish} \int V_G A = 0. \end{equation}
The quadratic operator contribution from the \(g_{o}\) term can be expressed as follows using the Schwinger representation of the string propagator:
\begin{equation}\label{oeff} \mathcal{O}_{\text{eff}}^{(g_{o})}(V_{G})=-g_{o}\int _{0}^{\infty }dt\int V_{G}\cdot b_{0}\left[e^{-tL_{0}}\left(\mathbb{I}-\mathbb{P}\right)A\ast A\right]. \end{equation}
The integrated stringy products on the right-hand side of~\eqref{effoperator} are formally evaluated using the standard definition of the \(2\)-string vertex:
\begin{equation}\label{nstringV} \int \Psi_1 \ast \Psi_2 = \left\langle h_1\cdot \Psi_1\left(0\right) h_2\cdot \Psi_2\left(0\right) \right\rangle_{\text{UHP}}, \end{equation}
where the conformal maps \(h_{k}\) map local coordinate patches of individual string states to the boundary of the UHP. These maps are given explicitly by:
\begin{equation}\label{fnk} h_1\left(z\right)=\frac{1+z}{1-z}, \qquad h_2\left(z\right)=-\frac{1-z}{1+z}. \end{equation}
The operator \(e^{-tL_{0}}\) is the generator of worldsheet dilatations (scaling transformations). A scaling transformation acting on the composite output state can be re-mapped via a conformal coordinate change directly onto the two input string fields \(\Psi _{1}\) and \(\Psi _{2}\). Under the state-operator correspondence, we define the \(t\)-dependent conformal maps \(h_1(z_1, t)\) and \(h_2(z_2, t)\) that project the local coordinate patches of the fields onto the boundary of the Upper Half-Plane (\(\mathbb{H}\)):
\begin{equation}\label{fnt} h_1\left(z,t\right)=e^{\frac{t}{2}}\left(\frac{1+z}{1-z}\right), \qquad h_2\left(z,t\right)=-e^{-\frac{t}{2}}\left(\frac{1-z}{1+z}\right). \end{equation}
Therefore, the effective operator~\eqref{oeff} is given by:
\begin{equation}\label{oeffev} \mathcal{O}_{\text{eff}}^{(g_{o})}(V_{G})=-g_{o}\int _{0}^{\infty }dt ~\oint \frac{dz}{2\pi i}\left\langle b(z) V_{G}(i,-i)\left(\mathbb{I}-\mathbb{P}\right)V_A\left( e^{\frac{t}{2}}\right)V_A\left( -e^{-\frac{t}{2}}\right)\right\rangle_{\text{UHP}}, \end{equation}
where \(\left(\mathbb{I}-\mathbb{P}\right)\) instructs us to remove the contributions to the worldsheet correlation function arising from the massless states flowing along the internal channels. The explicit \(t\)-integrated expression reduces directly to:
\begin{equation}\mathcal{O}_{\text{eff}}^{(g_{o})}[V_{G}]=\int d^{26}x\,h^{\mu \nu }(x)T_{\mu \nu }(x)+\mathcal{O}(\alpha ^{\prime }),\end{equation}
where \(T_{\mu \nu }\) is the exact Maxwell energy-momentum tensor.
This explicitly demonstrates that the \(g_{o}\) term in the effective operator captures the full quadratic component of the Maxwell energy-momentum tensor coupled to the background metric fluctuation.

\section{The emergence of the closed string vacuum from OSFT}\label{sec:Condensation}

A profound feature of Witten’s cubic OSFT is its capacity to describe physics beyond the perturbative open string spectrum. While the theory is formulated purely in terms of open string fields \(\Psi \), its non-perturbative dynamics naturally incorporate closed strings. This section reviews how the true closed string vacuum emerges as a specific classical solution to the OSFT equations of motion, a phenomenon deeply intertwined with Sen's conjectures and the mechanism of tachyon condensation \cite{Sen:1998ii, Sen:1998sm, Sen:1999nx, Berkovits:2000hf}. \par

\subsection{Tachyon condensation}\label{Tachcond}

\begin{figure}[t!]
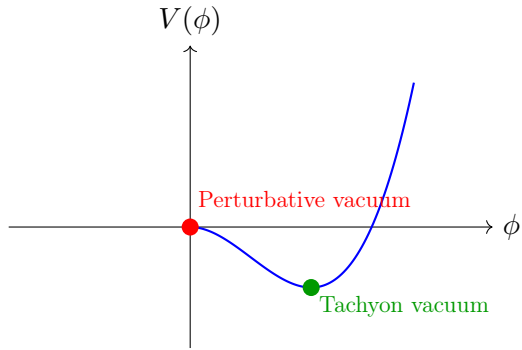

    \centering
    \potplot
    \caption{Tachyon condensation is the non-perturbative dynamical process wherein the classical tachyon field rolls down this potential from the unstable minimum toward a stable minimum. }
    \label{potplot}
\end{figure}

A space-filling \(D25\)-brane in the bosonic string theory or an unstable brane anti-brane pair in superstring theory possesses an open string spectrum whose lowest-lying state is a scalar field \(\phi(x)\) with a negative mass-squared \cite{Green:1994iw, Banks:1995ch,  Green:1996um, Lifschytz:1996iq}. The potential for the tachyonic field \(\phi\)  can be computed using the framework of Witten's cubic OSFT. Truncating the OSFT action \eqref{OSFTAction} to the level-0 tachyon mode \(\phi(x)\) gives the following approximation of the potential \cite{Sen:1999nx}
\begin{equation}\label{Tpotential}
V(\phi )=-\frac{1}{2\alpha ^{\prime }}\phi ^{2}+\frac{27\sqrt{3}}{16}g_o\phi ^{3}+\mathcal{O}(\phi ^{4})
\end{equation}
The negative quadratic coefficient confirms that the perturbative D-brane configuration sits at a local unstable maximum of the string field configuration space. Tachyon condensation is the non-perturbative dynamical process wherein the classical tachyon field rolls down this potential toward a global, stable minimum, see Figure \ref{potplot}.\par

\subsection{Sen's conjectures}\label{Senc}

 Sen's conjectures are three exact, non-perturbative statements describing the global structure of the open string tachyon potential \(V(\phi)\). These statements bridge the gap between abstract string field equations and target-space D-brane geometry.
 
 \underline{Conjecture 1}:  The difference in potential energy density between the unstable perturbative vacuum \( (\phi = 0)\) and the stable non-perturbative minimum \((\phi = \phi_0)\) exactly equals the total spatial tension \(T_{p}\) of the underlying Dp-brane system
 \begin{equation}
 V(0)-V(\phi _{0})=T_{p}
 \end{equation}
 At the global minimum  the negative potential energy precisely cancels the mass-energy density of the brane.

 \underline{Conjecture 2}:  While the global minimum represents a completely homogeneous vacuum, the classical string field equations  admit non-trivial, spatially dependent lump solutions (solitonic configurations). A lump solution \(\Psi_{\text{lump}}(x)\) that transitions asymptotically along a spatial direction \(x^i \to \infty\) to the stable minimum \(\phi_0\), but passes through the unstable maximum \(\phi=0\) at a localized codimensional interface, describes a stable lower-dimensional D(p-n)-brane, where \(n<p\).

 \underline{Conjecture 3}:  At the global stable minimum \(\Psi _{\text{tv}}\), the original D-brane has entirely dissolved. Because open strings require the boundary conditions provided by a physical D-brane to exist, there are no physical open string physical degrees of freedom left in the spectrum around \(\Psi _{\text{tv}}\).

Early evidence for Sen's conjectures relied exclusively on numerical level truncation, yielding highly precise yet non-rigorous approximations of the potential \cite{Sen:1999nx, Berkovits:2000hf}. This limitation was overcome by Schnabl \cite{Schnabl:2005gv}, who analytically proved Conjectures 1 and 3 by finding an exact classical solution for the tachyon vacuum, marking a crucial step toward understanding the closed string vacuum from an OSFT perspective.

\subsection{Wedge states}\label{wedge}

\begin{figure}[t!]
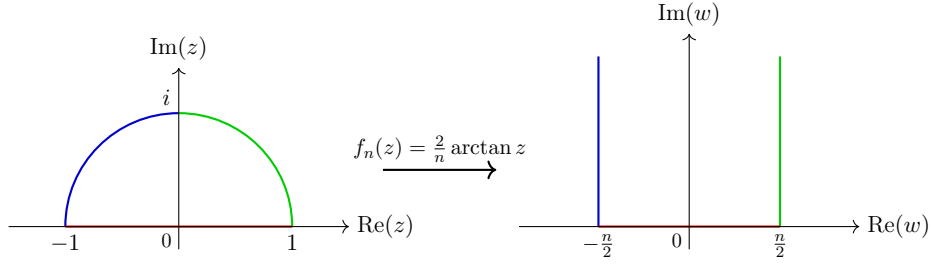

    \centering
    \wedgen
    \caption{The conformal map \(f_{n}(z) \) maps the unit half-disk to a semi-infinite strip of width \(n\). Under this transformation, the string midpoint \(z=i\) is mapped to \(w=i\infty\), while the left and right halves of the string are mapped to the right and left boundaries of the strip, respectively.}
    \label{wedgen}
\end{figure}

Schnabl constructed the analytic solution for tachyon condensation as an explicit Taylor-like series expansion in terms of wedge states \(\lvert \Omega^n \rangle\)  which can be defined as a linear functional over any test state \(\zeta \) in the CFT state-space
\begin{equation}\label{nphi}
\langle \Omega^n\mid \zeta \rangle =\left<f_{n}\circ \zeta (0)\right>_{\mathbb{H}}
\end{equation}
The explicit canonical conformal map that defines a wedge state of width \(n\) is given by
\begin{equation}\label{wedgemap}
f_{n}(z)=w= \frac{{2}}{{n}}\arctan {(z)}
\end{equation}
which maps a unit half-disk to a semi-infinite strip having width \(n\) \cite{Schnabl:2002gg}, see Figure \ref{wedgen}. The  mid-point of the string \(z=i\) is mapped to \(w=i\infty \) and the left and right halves of the string is mapped to the right and left edges of the strip.  The \(w\)-coordinate is usually called as the sliver frame.\par

Wedge states solved the greatest mathematical obstacle in open string field theory: the non-locality of Witten's star product. Computing the star product between two arbitrary oscillator states requires calculating an infinite matrix of complex Neumann coefficients \cite{Gross:1986ia}. However, because wedge states translate the star product into a simple addition of widths, they turn difficult non-local worldsheet path integrals into manageable, exact algebra.\par

 The standard perturbative open string \(SL(2,\mathbb{R})\)  vacuum state \(\lvert \Omega \rangle\)  corresponds geometrically to a semi-infinite strip of width \(1 \). A wedge state of width \(n\) is constructed by taking the standard vacuum state \(\lvert \Omega \rangle\) and repeatedly multiplying it by itself using Witten's non-local midpoint star product 
\begin{equation}\label{nfrom2}
|\Omega^n\rangle =\underbrace{{|\Omega\rangle \ast |\Omega\rangle \ast \dots \ast |\Omega\rangle }}_{{n}\text{ times}}
\end{equation}
This definition naturally satisfies a very powerful and simple algebraic fusion rule
\begin{equation}\label{fusion}
{|\Omega^n\rangle \star |\Omega^m\rangle =|\Omega^{n+m}\rangle }
\end{equation}
This structural simplicity is what ultimately allowed Schnabl to write down exact solutions for the tachyon vacuum. \par

Interestingly, the wedge state \(|\Omega^n\rangle\) satisfy the following differential equation \cite{Erler:2019vhl, Okawa:2006vm}
\begin{equation}\label{modK}
\frac{d}{dn}|\Omega^n\rangle=-n |K\rangle\ast |\Omega^n\rangle
\end{equation}
\(K\) is the infinitely long thin vertical strips carrying the operator insertion \(\mathcal{K}\)
\begin{equation}\label{L}
 \mathcal{K}=\int_{-i\infty}^{i\infty}\frac{dw}{2\pi i}~ T(w)
\end{equation}

\subsection{Schnabl's  solution for tachyon condensation}\label{schnabl}

The Schnabl's analytic solution for the tachyon vacuum \cite{Schnabl:2005gv} can be constructed by  recursively solving the  OSFT equation of motion in the Schnabl gauge 
\begin{equation}\label{B}
\mathcal{B}\Psi =0,\qquad \qquad  \mathcal{B}=\int_{-i\infty}^{i\infty}\frac{dw}{2\pi i}~ b(w)
\end{equation}
where \(w\) is the coordinate on the infinitely long thin strip. The solution consists of two pieces and is defined as the following limit
\begin{equation}\label{schnablsol}
\Psi_{Sch}=\lim_{N\to \infty}\left[\psi_N -\sum_{n=0}^N\psi_n\right]
\end{equation}
The state \(\psi_n\) is given by
\begin{equation}\label{psin}
\psi_n= |\Omega^{\frac{1}{2}}\rangle \ast c\ast B \ast |\Omega^n\rangle\ast c\ast |\Omega^{\frac{1}{2}}\rangle
\end{equation}
The string field \(c\) is defined by an infinitely long  thin strip  with a boundary insertion (on the real axis) of the \(c(w)\)-ghost and \(B\) is the infinitely long thin vertical strips carrying the operator insertion \(\mathcal{B}\).\par

Schnabl’s solution provided the first mathematically exact, non-perturbative proof of Sen’s Conjectures. It analytically proved that the vacuum potential energy precisely cancels D-brane tension. Moreover, it definitively demonstrated that the open string BRST cohomology becomes entirely empty in the tachyon vacuum.

\subsection{\(KBc\) algebra and closed form real analytic solution}

Schnabl’s original solution is written as a delicate, infinite series expansion over wedge states. Manipulating this solution requires keeping careful track of infinite limits, regularizing worldsheet divergence patterns, and handling a mathematically sensitive boundary anomaly term. This makes computing even basic scattering amplitudes incredibly tedious. \par

However,  Okawa realized that Schnabl's series could be recast using three simple string field operators that form a closed associative algebra under the Witten star product: \(K, B,\) and \(c\) \cite{Okawa:2006vm}. The defining relations of the \(KBc\) algebra are
\begin{equation}\label{KBcalgebra}
K\ast B-B\ast K=0 \qquad B\ast c+c\ast B=1\qquad B\ast B=0 \qquad  c\ast c=0
\end{equation}
The BRST charge \(Q_{B}\) acts as a differential operator on the algebra. It treats the fields as a differential graded algebra via the following derivation rules
\begin{equation}\label{BRSTonKBc}
Q_{B}K=0 \qquad Q_{B}B=K \qquad Q_{B}c=c\ast K\ast c
\end{equation}

To make Schnabl's solution more algebraically transparent, Yuji Okawa reformulated it into a beautifully compact, closed-form geometric expression utilizing the \(KBc\)-algebra
\begin{equation}\label{okawa}
\Psi_{Oka} =e^{-K/2}\ast c\ast \frac{K}{1-e^{-K}}\ast B \ast c\ast e^{-K/2}
\end{equation}
The Schnabl's series expression can be recovered by expanding the algebraic denominator as a geometric series and mapping the abstract operator \(K\) to its geometric worldsheet counterparts. \par

While Schnabl’s original solution and Okawa's subsequent parameterization successfully proved Sen's conjecture, they were not real string fields under the standard OSFT reality condition. Erler and Schnabl solved this by constructing a beautifully simple, geometrically well-behaved, and manifestly real closed-form solution within the \(KBc\) algebra \cite{Erler:2009uj}, which is defined compactly as
\begin{equation}\label{EStvac}
\Psi _{\text{ES}}=\frac{1}{\sqrt{1+K}}\ast c\ast\frac{1+K}{K}\ast B\ast c\ast \frac{1}{\sqrt{1+K}}
\end{equation}
This string field is "real," meaning it is invariant under a combination of worldsheet parity (reversing the string coordinates \(\sigma \to \pi - \sigma\)) and complex conjugation. Moreover, it is strictly analytic and regular for all positive values of (\(K \geq 0\)).  We will demonstrate in the following section that these features make Erler and Schnabl  solution well suited for computing correlation functions of the gauge invariant OSFT operators around the tachyon vacuum. 

\section{Perturbation theory around the tachyon vacuum}\label{sec:Perturbation}

In this section, we develop the perturbation theory around the tachyon vacuum, which serves as the foundation for studying the correlation functions of gauge-invariant operators in the subsequent section.

\subsection{The modified BRST operator and the vanishing cohomology}\label{ModQB}

 To perform perturbation theory around the tachyon vacuum, let us expand the full string field \(\Psi \) around the classical tachyon vacuum solution \(\Psi _{tv}\) such as the Erler-Schnabl or Okawa solution: 
 \begin{equation}\nonumber\\
 \Psi =\Psi _{tv}+\chi 
 \end{equation}
 where \(\chi \) represents the fluctuation. Substituting into the Witten cubic action gives again a cubic action for \(\chi \) 
 \begin{equation}\label{chiaction}
 S[\Psi ]=S[\Psi_{tv} ]+\int \left(\frac{1}{2}\chi *Q_{\Psi _{tv}}\chi +\frac{g_o}{3}\chi *\chi *\chi \right)
 \end{equation}
 The linear term vanishes by the equation of motion, and the kinetic term for the fluctuation \(\chi \) shifts to a new background-dependent BRST operator
\begin{equation}\label{ModQB}
Q_{\Psi _{tv}}\chi =Q_{B}\chi +\Psi _{tv}*\chi -(-1)^{|{}\chi |{}}\chi *\Psi _{tv}
\end{equation}

In standard perturbation theory, physical asymptotic particle states correspond to the cohomology of the BRST operator. Around the tachyon vacuum, however, the cohomology of \(Q_{\Psi _{tv}}\) is completely empty. This is proven via the existence of the homotopy operator \(A\), which satisfies
\begin{equation}\label{Hop}
\{Q_{\Psi _{tv}},A\}=\mathbf{1}
\end{equation}
Consequently, any physical on-shell open string fluctuation \(\chi \) defined such that \(Q_{\Psi_{tv}} \chi = 0\), it is automatically exact (pure gauge)
\begin{equation}\label{exact}
\chi =\mathbf{1}*\chi =(Q_{\Psi _{tv}}A+AQ_{\Psi _{tv}})\chi =Q_{\Psi _{tv}}(A*\chi )
\end{equation}
Because every closed state is exact, there are no physical perturbative open string states. The tachyon, the gauge field (photon), and all higher-spin string oscillations are entirely removed from the spectrum.

\subsection{Feynman rules around the tachyon vacuum}\label{Mprop}

The perturbative expansion of time-ordered correlation functions in a quantum field theory can be mapped onto a diagrammatic calculus using Feynman rules, where lines represent field contractions (propagators) and intersections denote interaction vertices. \par

The Feynman rules for OSFT around the tachyon vacuum are identified by expanding the string field fluctuation \(\lvert\chi\rangle\) in the oscillator basis of the worldsheet Hilbert space:
\begin{equation}\label{chif} 
\vert{}\chi\rangle=\sum_{\alpha}\int \frac{d^{26}k}{\left(2\pi\right)^{26}}~ \chi_{\alpha}\left(k\right)\vert{}k,\alpha\rangle,  
\end{equation}
where \(\{\chi_{\alpha}\left(k\right)\}\) denote the Fourier modes of the target spacetime fields. The kinetic operator \(\mathcal{Q}_{\beta \alpha}(k)\) is given by
\begin{equation}
\mathcal{Q}_{\beta \alpha }(k)\delta ^{(26)}\left(k-{\tilde k}\right)\equiv \langle {\tilde k},\beta |{}Q_{\Psi _{tv}}|{}k,\alpha \rangle ,
\end{equation}
while the 3-string vertex factor is defined as
\begin{equation}\label{vertex}  
V_{\beta \beta_1 \beta_2}(k, {\tilde k}_1, {\tilde k}_2) \equiv \langle k, \beta \vert{}\left(\vert{}{\tilde k}_1, \beta_1\rangle * \vert{}{\tilde k}_2, \beta_2\rangle\right).
  \end{equation}
  
  To construct the perturbation theory, one must invert the kinetic operator. This inversion is achieved by exploiting a defining algebraic feature of the tachyon vacuum background: the existence of a homotopy operator \(A\). For the Erler-Schnabl tachyon solution \(\Psi _{ES}\) \eqref{EStvac}, the homotopy operator takes the form
  \begin{equation}\label{AES}  
  A =B\ast  \frac{1}{1+{K}}=B\ast\int_{0}^{\infty}e^{-t}~\Omega^t.  
  \end{equation}
where we used the fact that
  \begin{equation}\label{OmK}  
  e^{-K}=\Omega, 
  \end{equation}
   which is a direct consequence of the relation \eqref{modK}. Remarkably, within Schnabl gauge, the  string propagator \(\Pi \) is identified precisely with this homotopy operator. \par
  
The perturbative expansion of OSFT correlation functions can be constructed from trivalent  graphs using the following Feynman rules:
\begin{itemize}

\item \textbf{Propagator (Edges):} For an internal edge \(e_{ij}\) connecting indices \(\beta_i\) and \(\beta_j\) with momentum \(q_{ij}\), assign the propagator:
\begin{equation}
\Pi_{\beta_i \beta_j}(q_{ij}) = \mathcal{Q}^{-1}_{\beta_i \beta_j}(q_{ij}).
\end{equation}

\item \textbf{Interaction (Vertices):} For a trivalent vertex \(v\) joining three strings with incoming momenta \(q_v^{1}, q_v^{2}, q_v^{3}\), assign the vertex factor:
\begin{equation}
-g_{o} \cdot \delta^{(26)}(q_v^{1} + q_v^{2} + q_v^{3}) \, V_{\beta_v^{1} \beta_v^{2} \beta_v^3}(q_v^{1}, q_v^{2}, q_v^{3}).
\end{equation}

\item \textbf{Loop Integrals:} For each independent internal loop momentum \(p_r\) unconstrained by momentum conservation, assign the integration measure:
\begin{equation}
\int \frac{d^{26}p_r}{(2\pi)^{26}}.
\end{equation}

\end{itemize}

  Using the Feynman rules we can expand the correlation function as a power series in the string coupling \(g_{o}\):
\begin{equation}\label{Sexpansion}
\langle \!\langle \chi_{\alpha_1}\left(k_1\right) \cdots \chi_{\alpha_n}\left(k_n\right) \rangle \!\rangle =\sum_{m=0}^{\infty}g_o^m\langle \!\langle \chi_{\alpha_1}\left(k_1\right) \cdots \chi_{\alpha_n}\left(k_n\right) \rangle \!\rangle ^{(m)}
\end{equation}
The \(l\)-loop contribution to the general \(n\)-point connected correlation function can be written compactly as:
\begin{align}\label{nlcorr}
 &\langle \!\langle \chi_{\alpha_1}\left(k_1\right) \cdots \chi_{\alpha_n}\left(k_n\right) \rangle \!\rangle ^{(2l+n-2)}\nonumber\\ 
 &=\sum_{\left\{\Gamma^n_{l}\right\}}\sum_{\{\beta\}}\frac{1}{\left\vert{}\text{Sym}\left(\Gamma^n_{l}\right)\right\vert{}}\int \prod _{r=1}^{l}\left[\frac{d^{26}p_{r }}{(2\pi )^{26}}\right]\prod _{ \{E_{ij}\}}\Pi _{\alpha_i \beta_{j}}(k_{i}) \prod _{\{e_{ij}\}}\Pi _{\beta_i \beta_{j}}(q_{ij})\prod _{\{v\}}V_{\beta_{v}^{1}\beta_{v}^{2}\beta_{v}^{3}}(q_{v}^{1},q_{v}^{2}, q_{v}^{3}), 
 \end{align}
 where \(\left\{\Gamma _{l}^{n}\right\}\) is the set of all trivalent  graphs with \(n\) external lines and \(l\) loops. The sets \(\{E_{ij}\}\), \(\{e_{ij}\}\), and \(\{v\}\) denote the external legs, internal edges, and trivalent vertices of a given graph \(\Gamma _{l}^{n}\), respectively. The topological parameters, namely the number of vertices \(V\) and the number of internal propagators \(E\), are strictly constrained by the graph topology:
 \begin{equation}\label{EVnumber}
   V = 2l + n - 2, \qquad E = 3l + n - 3. 
  \end{equation}
  By enforcing momentum conservation at each vertex, the internal line momenta \(q_{ij}\) are uniquely determined as linear combinations of the external momenta \(k_{i}\) and the unconstrained virtual loop momenta \(p_{r}\). 
  
\subsection{The worldsheet moduli space representation}\label{Wsheet}

\begin{figure}[t!]
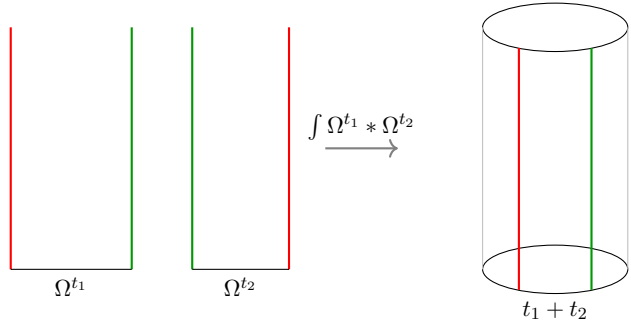

    \centering
   \innerp
    \caption{The star product of two wedge states, \(\lvert \Omega^{t_1} \rangle\) and \(\lvert \Omega^{t_2} \rangle\), followed by integration, yields a semi-infinite cylinder with a circumference of \(t_1+t_2\).}
    \label{innerp}
\end{figure}

The transition from the algebraic expression of OSFT correlation functions to their worldsheet moduli space representation requires a geometric description of the underlying propagators and vertices. To facilitate this mapping, it is convenient to define a modified propagator \(\hat{\Pi }_{\alpha \beta }(k,\tilde{k})\) by augmenting the standard propagator \(\Pi_{\alpha\beta}(k)\) with a momentum delta function
  \begin{equation}\label{prop}
  \hat \Pi_{ \alpha\beta}(k,\tilde k)  = \langle {  k}, \alpha \vert{} A \vert{}{ \tilde k}, \beta \rangle
  \end{equation}
 The geometric worldsheet description of the propagators and vertices can be identified by specializing to a choice of tachyon vacuum background. For our purpose, the most convenient selection is the Erler-Schnabl solution. Utilizing the explicit form of its associated homotopy operator \(A\) \eqref{AES} alongside the state-operator correspondence \cite{Erler:2019vhl},
 \begin{equation}\label{vertstate}
  \lvert k,\alpha\rangle \ \to \ \Omega^{\frac{1}{2}} \mathcal{V}_{k}^{\alpha}\Omega^{\frac{1}{2}},
 \end{equation}
 we can recast the propagator in terms of wedge states as follows:
      \begin{equation}\label{propw}
  \hat \Pi_{ \alpha \beta}(k,\tilde k) =\int_{0}^{\infty}dt ~e^{-t} \int \Omega^{\frac{1}{2}} \mathcal{V}_{k}^{\alpha}\Omega^{\frac{1}{2}}\ast B\ast \Omega^{t}\ast  \Omega^{\frac{1}{2}} \mathcal{V}_{\tilde k}^{\beta}\Omega^{\frac{1}{2}}
  \end{equation}
 where the second integration denotes the formal Witten's open string trace.

As discussed in Section \ref{wedge}, the star product of two wedge states, \(\lvert \Omega^{t_1} \rangle\) and \(\lvert \Omega^{t_2} \rangle\), geometrically glues their associated strips side-by-side. Formally integrating the resulting state connects the far-left and far-right boundaries of this combined strip of width \(t_1+t_2\). This topological closure wraps the flat domain into a semi-infinite cylinder of circumference \(t_1+t_2\) (see Figure \ref{innerp}). Consequently, the sequential operation of the star product and the open string trace can be identified with the BPZ inner product \(\langle \Omega^{t_1} \vert{} \Omega^{t_2} \rangle\). \par

\begin{figure}[t!]
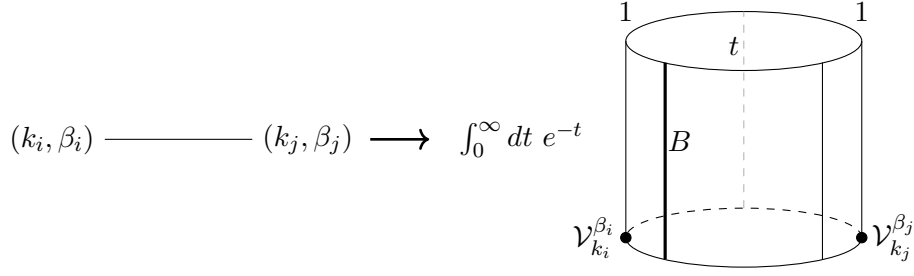

    \centering
    \prop
   \caption{Worldsheet geometry of the propagator \(\hat \Pi_{\alpha\beta}(k,\tilde k)\). The configuration maps onto a semi-infinite cylinder of circumference \(t+2\), with the ghost line operator \(B\) integrated vertically along its length. The open string vertex operators are inserted at the boundary base, separated by a unit distance.}   \label{propa} 
\end{figure}

Therefore, the propagator \eqref{propw} maps onto a correlation function of the boundary vertex operators \(\mathcal{V}_{k}^{\alpha}\) and \(\mathcal{V}_{\tilde k}^{\beta}\) evaluated on a semi-infinite cylinder of circumference \(t+2\). In this geometric picture, as described in Figure \ref{propa}, the line-like ghost operator \(B\) \eqref{B} is inserted along an infinitely long vertical line, and the modulus \(t\)—representing the Schwinger parameter—is integrated from \(0\) to \(\infty\). The open string vertex operators are pinned to the bottom boundary of the cylinder, separated by a unit distance. \par

\begin{figure}[t!]
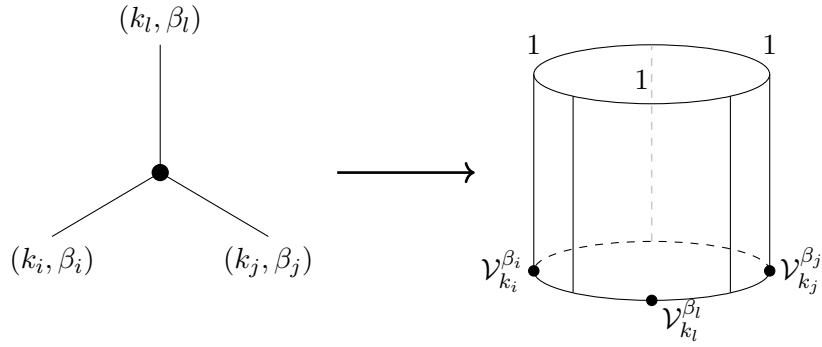

    \centering
    \cubicv
    \caption{The three open string fields fuse to generate a semi-infinite cylinder of fixed circumference \(3\). The three vertex operators are sequentially inserted along the bottom boundary with a uniform spacing of unit length.}
    \label{cubicv}
\end{figure}

Similarly, the cubic vertex \eqref{vertex} admits a direct worldsheet path integral representation:
\begin{equation}\label{vertexw}
V_{\beta_i\beta_j\beta_l}(k_i,k_j,k_l)=\int \left( \Omega^{\frac{1}{2}} \mathcal{V}_{k_i}^{\beta_i}\Omega^{\frac{1}{2}} \right) \ast \left( \Omega^{\frac{1}{2}} \mathcal{V}_{k_j}^{\beta_j}\Omega^{\frac{1}{2}} \right) \ast \left( \Omega^{\frac{1}{2}} \mathcal{V}_{ k_l}^{\beta_l}\Omega^{\frac{1}{2}} \right).
\end{equation}
Geometrically, as shown in Figure \ref{cubicv}, this corresponds to a correlation function of \(\mathcal{V}_{k_i}^{\beta_i}\), \(\mathcal{V}_{ k_j}^{\beta_j}\), and \(\mathcal{V}_{ k_l}^{\beta_l}\) on a fixed semi-infinite cylinder of circumference \(3\). The three vertex operators are inserted along the base boundary, separated from each other by a uniform unit distance. \par

\begin{figure}[t!]
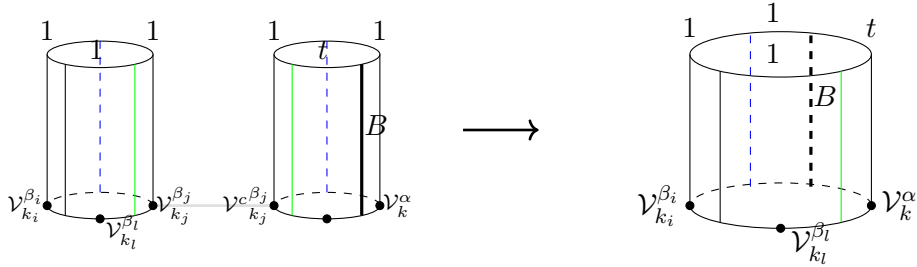

    \centering
   \ProptVA
    \caption{The worldsheets associated with the propagator \(\hat{\Pi }_{\beta _{j}\alpha }(k_{j},k)\) and the cubic vertex \(V_{\beta_i\beta_j\beta_l}(k_i,k_j,k_l)\) are glued by summing over the continuous and discrete labels \(k_{j}\) and \(\beta _{j}\). This operation eliminates the unit-width vertical strips containing the vertex operators \(\mathcal{V}_{k_{j}}^{\beta _{j}}\) and \(\mathcal{V}_{k_{j}}^{c\beta _{j}}\), sewing the remaining domains along the boundaries denoted by the green and blue vertical lines. The resulting surface is a semi-infinite cylinder of circumference \(t+3\) featuring a line-like \(B\)-ghost insertion, with the vertex operators \(\mathcal{V}_{k_{i}}^{\beta _{i}}\), \(\mathcal{V}_{k_{l}}^{\beta _{l}}\), and \(\mathcal{V}_{k}^{\alpha }\) anchored at its base.}
    \label{ProptVA}
\end{figure}

Arbitrary off-shell correlation functions are systematically constructed on the worldsheet by sewing the geometric surfaces of propagators onto those of cubic vertices. To illuminate this gluing mechanism, consider the propagator worldsheet \(\hat{\Pi}_{\beta_{j}\alpha}(k_{j},k)\) (Figure \ref{propa}) and the cubic vertex worldsheet \(V_{\beta_i\beta_j\beta_l}(k_i,k_j,k_l)\) (Figure \ref{cubicv}). As illustrated in Figure \ref{ProptVA}, the vertex operator \(\mathcal{V}_{k_{j}}^{\beta _{j}}\) generates a state along the green and blue vertical boundaries of a unit-width, semi-infinite strip on the vertex cylinder, while its conjugate operator on the propagator cylinder inserts the dual state along the corresponding colored lines. According to the sewing and smoothing theorem \cite{Sonoda:1988mf,Sonoda:1988fq,Schwarz:1989qx}, summing over a complete set of intermediate states in the worldsheet Hilbert space effectively excises these unit-width strips from both cylinders and smoothly joins the remaining surfaces along their matching boundaries. The resulting Riemann surface is a semi-infinite cylinder of circumference \(t+3\), featuring a line-like \(B\)-ghost insertion, with the vertex operators \(\mathcal{V}_{k_{i}}^{\beta _{i}}\), \(\mathcal{V}_{k_{l}}^{\beta _{l}}\), and \(\mathcal{V}_{k}^{\alpha }\) securely anchored at its base. \par

\begin{figure}[t!]
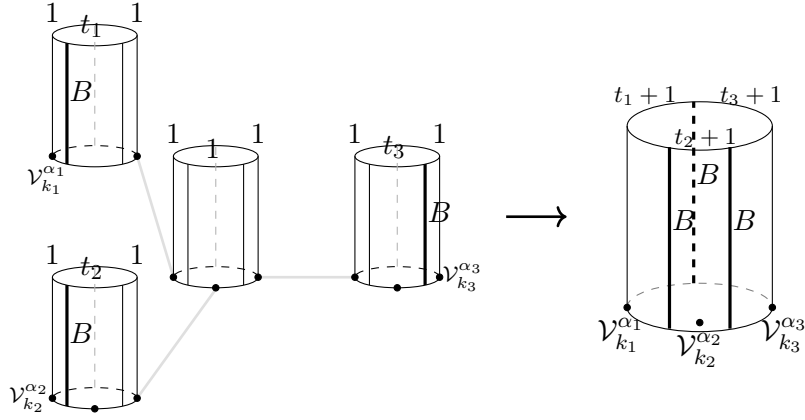

    \centering
   \ProptV
    \caption{Worldsheet geometry of the off-shell three-point correlation function. The resulting smooth Riemann surface is a semi-infinite cylinder of total circumference \(\sum_{i=1}^3 t_i + 3\), formed by sequentially sewing three propagators onto the cubic vertex. The worldsheet features three separate vertical line-like \(B\)-ghost insertions, with the external open string vertex operators \(\mathcal{V}_{k_{1}}^{\alpha _{1}}\), \(\mathcal{V}_{k_{2}}^{\alpha _{2}}\), and \(\mathcal{V}_{k_{3}}^{\alpha _{3}}\) securely anchored along the base boundary.}
    \label{ProptV}
\end{figure}

Applying this gluing prescription, we can straightforwardly derive the worldsheet moduli space expression for the three-point off-shell correlation function.
\begin{align}\label{3lcorr}
\langle \!\langle \chi_{\alpha_1}\left(k_1\right)  \chi_{\alpha_2}\left(k_2\right)  \chi_{\alpha_3}\left(k_3\right) \rangle \!\rangle ^{(1)}&\nonumber\\
=\sum_{\left\{\beta_{v}^j\right\}}\int \prod_{i=1}^{3}\frac{d^{26}q_{v^i}}{\left(2\pi\right)^{26}}&\hat\Pi _{\alpha_1 \beta^1_{v}}(k_i,q^1_{v})\hat\Pi _{\alpha_2 \beta^2_{v}}(k_i,q^2_{v})\hat\Pi _{\alpha_3 \beta^1_{v}}(k_i,q^1_{v})  V_{\beta_{v}^{1}\beta_{v}^{2}\beta_{v}^{3}}(q_{v}^{1},q_{v}^{2}, q_{v}^{3})
\end{align}
The associated worldsheet is an semi-infinite cylinder of circumference \(\sum_{i=1}^3t_i+3\), featuring three \(B\)-ghosts insertions with the vertex operators \(\mathcal{V}_{k_{1}}^{\alpha_{1}}\), \(\mathcal{V}_{k_{2}}^{\alpha_{2}}\), and \(\mathcal{V}_{k_3}^{\alpha_3 }\) anchored at its base, see  Figure \ref{ProptV}. \par

\begin{figure}[t!]
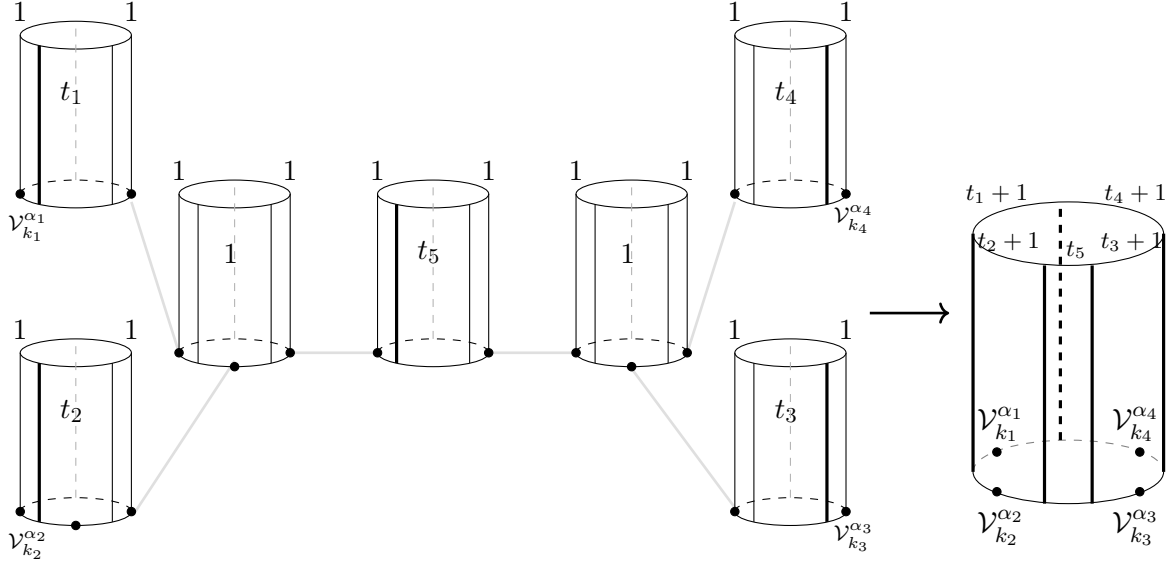

    \centering
   \schannG
    \caption{The \(s\)-channel 4-point worldsheet configuration. A semi-infinite cylinder of circumference \(\sum_{i=1}^5 t_i + 4\) generated by sewing five propagators and two trivalent vertices. The four external vertex operators are pinned to the base boundary, and the five thick lines represent the integrated \(B\)-ghost insertions.}
    \label{schannG}
\end{figure}

A general tree-level \(n\)-point correlator computed using the gluing prescription takes the form of an integral over \((2n-3)\) Schwinger parameters:
\begin{align}\label{nTreecorr}
\langle \!\langle \chi_{\alpha_1}\left(k_1\right)  \cdots  \chi_{\alpha_n}\left(k_n\right) \rangle \!\rangle ^{(2n-3)}=\sum_{\Gamma^n_0}\int_{0}^{\infty} \prod_{i=1}^{2n-3}dt_i(\Gamma^n_0) \, e^{-t_i(\Gamma^n_0)}\,\langle \mathcal{V}_{k_1}^{\alpha_1}\cdots\mathcal{V}_{k_n}^{\alpha_n}\mathcal{B}(\Gamma^n_0)\rangle_{C(\Gamma^n_0)},
\end{align} 
where \(\Gamma^n_0\) represents an arbitrary trivalent tree graph with \(n\) external legs, and \(\mathcal{B}(\Gamma^n_0)\) denotes the composite \(b\)-ghost insertion for all propagators in the graph. The sum runs over all inequivalent tree-level trivalent diagrams contributing to the correlator. The worldsheet integrand is evaluated on \(C(\Gamma^n_0)\), a semi-infinite cylinder of circumference \(\left(\sum_{i=1}^{2n-3} t_i(\Gamma^n_0)+n\right)\), which is capped by a hemisphere at infinity. The positions of the vertex operators anchored at the base of the cylinder are determined by the Schwinger parameters and the topology of the graph \(\Gamma^n_0\). The geometric construction of  \(C(\Gamma^4_0)\) associated with the \(s\)-channel graph that contributes to the 4-point correlation function is illustrated  in Figure \ref{schannG}. \par 

The expression \eqref{nTreecorr} can be straightforwardly generalized to all orders in perturbation theory. For topological clarity, we characterize the underlying graphs by their genus \(g\) and number of boundaries \(b\) rather than their graph-theoretic loop count \(l\). This classification is mediated by the topological constraint:
\begin{equation}\label{fatgraph}
2g+b=l+1.
\end{equation}
An arbitrary \(n\)-point correlator can then be expressed as a joint expansion over the genus and the number of boundaries:
\begin{align}\label{nallcorr}
&\langle \!\langle \chi_{\alpha_1}\left(k_1\right)  \cdots  \chi_{\alpha_n}\left(k_n\right) \rangle \!\rangle \nonumber\\
&\qquad=\sum_{g=0}^{\infty}\sum_{b=1}^{\infty}\sum_{\Gamma^n_{g,b}}g_0^{-2\chi_{g,n,b}}\int_{0}^{\infty} \prod_{i=1}^{d_{g,n,b}}dt_i(\Gamma^n_{g,b}) \, e^{-t_i(\Gamma^n_{g,b})}\,\langle \mathcal{V}_{k_1}^{\alpha_1}\cdots\mathcal{V}_{k_n}^{\alpha_n} \mathcal{B}(\Gamma^n_{g,b})\rangle_{C(\Gamma^n_{g,b})}.
\end{align} 
where \(\mathcal{B}(\Gamma^n_{g,b})\) is the \(b\)-ghost insertion associated with the trivalent graph \(\Gamma^n_{g,b}\), which possesses \(d_{g,n,b}\) propagators. The number of propagators is related to the genus, number of external legs and the boundaries as follows
\begin{equation}\label{dgnb}
d_{g,n,b} = 6g + 2n + 3b - 6.
\end{equation}
 The worldsheet correlator is evaluated on a Riemann surface \(C(\Gamma^n_{g,b})\) with Euler characteristic 
\begin{equation}\label{EChar}
\chi_{g,n,b}=2-2g-n-b.
\end{equation}
Geometrically, this surface consists of \(b\) semi-infinite cylinders capped at infinity by a genus-\(g\) surface with \(b\) boundaries. The total combined circumference of the cylinders equals \(\left(\sum_{i=1}^{d_{g,n,b}}t_i(\Gamma^n_{g,b})+n\right)\). \par
 
We conclude this section by emphasizing a definitive feature of this framework: whereas an infinitely long string strip (\(t \to \infty\)) inside a standard open string loop introduces severe infrared (IR) divergences around the perturbative vacuum—driven by the propagation of massless gauge fields or open string tachyons—the situation alters radically around the tachyon vacuum. Here, the bare propagator naturally supplies a mandatory exponential damping factor \(e^{-t}\) for every internal strip. This forces the worldsheet integrand to decay exponentially at the boundaries of moduli space, structurally rendering the full amplitude entirely free of IR divergences originating from on-shell open string excitations. 

\section{Closed string amplitudes from OSFT}\label{sec:Smatrix}
 
Around the tachyon vacuum, all on-shell open string states become exact with respect to the modified BRST operator. Consequently,   the on-shell limit of the arbitrary open string correlation functions constructed in the previous section vanishes identically.  Nevertheless, these off-shell correlation functions remain highly non-trivial and serve as the core building blocks for constructing the correlators of gauge-invariant observables in OSFT, which capture the true closed string dynamics.

 \subsection{Correlation functions of the gauge invariant operators}\label{CorrGinv}

\begin{figure}[t!]
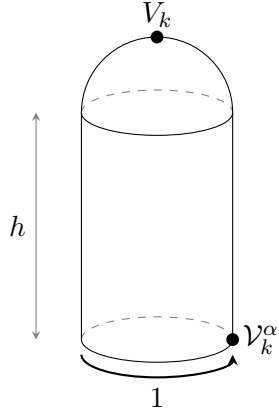

\centering
\GinvC
\caption{Worldsheet geometry of the gauge-invariant observable \(\mathcal{O}_{V}\). The operator is geometrically defined on a semi-infinite cylinder (\(h\to \infty\)) of unit circumference, capped asymptotically by a hemisphere at infinity. An OSFT boundary vertex operator \(\mathcal{V}_{k}^{\alpha }\) is anchored at the base of the cylinder, while the on-shell bulk vertex operator \(V_{k}\) is inserted at the north pole of the capping hemisphere.}     \label{GinvC} \end{figure}
In the sliver frame, the gauge-invariant observable \(\mathcal{O}_{V}\) is geometrically defined on a semi-infinite cylinder of unit circumference, asymptotically capped by a hemisphere at infinity. In this setup, an OSFT boundary vertex operator \(\mathcal{V}\) is anchored at the base of the cylinder, while an on-shell bulk vertex operator \(V\) is inserted at the north pole of the capping hemisphere (see Figure \ref{GinvC}). The correlation function of \(n\) such gauge-invariant operators around the tachyon vacuum is governed by the OSFT Feynman rules and takes the form:
\begin{align}\label{nSmatrix} \langle \!\langle &\mathcal{O}_{V_1}\left(k_1\right)  \cdots  \mathcal{O}_{V_n}\left(k_n\right)  \rangle \!\rangle \nonumber\\ &=\sum_{g=0}^{\infty}\sum_{b=1}^{\infty}\sum_{\Gamma^n_{g,b}}\sum_{\{\alpha_j\}}\int_{0}^{\infty} \prod_{i=1}^{d_{g,n,b}}dt_i({\Gamma^n_{g,b}}) e^{-t_i({\Gamma^n_{g,b}})}\langle\mathcal{V}_{k_1}^{\alpha_1}\cdots\mathcal{V}_{k_n}^{\alpha_n} \mathcal{B}(\Gamma^n_{g,b})\rangle_{C(\Gamma^n_{g,b})}\prod_{i=1}^n\langle V^i_{k_i}(i\infty){\hat{\mathcal{V}}}^{\alpha_i}_{k_i}\rangle_{C_1}\nonumber\\ &=\sum_{g=0}^{\infty}\sum_{b=1}^{\infty}\sum_{\Gamma^n_{g,b}}\int_{0}^{\infty} \prod_{i=1}^{d_{g,n,b}}dt_i({\Gamma^n_{g,b}})  e^{-t_i({\Gamma^n_{g,b}})}\langle{V}_{k_1}^1(z_1)\cdots{V}^n_{k_n}(z_n)\mathcal{B}(\Gamma^n_{g,b})\rangle_{\Sigma(\Gamma^n_{g,b})}  \end{align}
where \(\hat{\mathcal{V}}_{k_{i}}^{\alpha _{i}}\) is the BPZ conjugate of the boundary vertex operator \(\mathcal{V}_{k_{i}}^{\alpha _{i}}\). The transition from the second to the third line is achieved by interpreting the summation over all basis elements of the open string worldsheet Hilbert space—characterized by the index set \(\{\alpha_j\}\)—as the sewing of worldsheet path integrals. This gluing is implemented by sewing \(n\) semi-infinite cylinders of unit circumference, denoted as \(C_{1}\), to the worldsheet \(C({\Gamma^n _{g,b}})\) that represents the graph \(\Gamma _{g,b}^{n}\) contribution (with genus \(g\) and \(b\) boundaries) to the correlation function \(\langle \!\langle \chi_{\alpha_1}\left(k_1\right) \cdots \chi_{\alpha_n}\left(k_n\right) \rangle \!\rangle\).
\begin{figure}[t!]
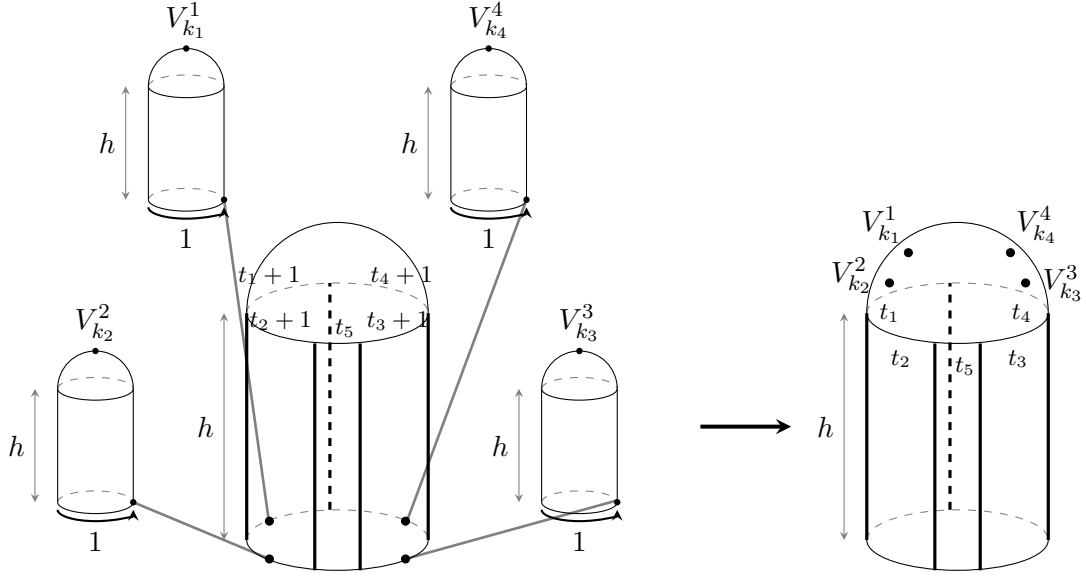

\centering
\Gincorr
\caption{The sewing procedure removes the interior of a unit strip from each unit-circumference cylinder along with the corresponding four unit strips from the boundary regions of \(C({\Gamma^4 _{0,1}})\) where the boundary vertex operators are anchored, followed by the identification of the respective boundaries. The resulting semi-infinite cylinder with unit circumference capped at infinity by a hemisphere with four bulk vertex operator insertions is obtained in the limit \(h\to\infty\).}     \label{Gincorr} \end{figure}
The sewing procedure removes the interior of a unit strip from each \(C_{1}\) cylinder and a corresponding unit strip from the neighborhood of the \(n\) marked points of \(C({\Gamma^n _{g,b}})\) where the boundary vertex operators are anchored, followed by the identification of the respective boundaries. The resulting surface, denoted as \(\Sigma(\Gamma^n_{g,b})\), is a Riemann surface of genus \(g\) with \(b\) boundaries. Geometrically, it can be visualized as \(b\) semi-infinite cylinders capped at infinity by a genus-\(g\) surface with \(b\) boundaries, where the total circumference of these cylinders sums to \(\sum_{i=1}^{d_{g,n,b}} t_i({\Gamma^n_{g,b}})\). Moreover, the genus-\(g\) cap at infinity carries the on-shell bulk vertex operators \(V_{k_1}^1, \dots, V_{k_n}^n\). This sewing construction is illustrated in Figure \ref{Gincorr} for the case \(g=0\), \(b=1\), and \(n=4\).

\subsection{Closing the worldsheet boundaries}\label{CloseWb}

\begin{figure}[t!]
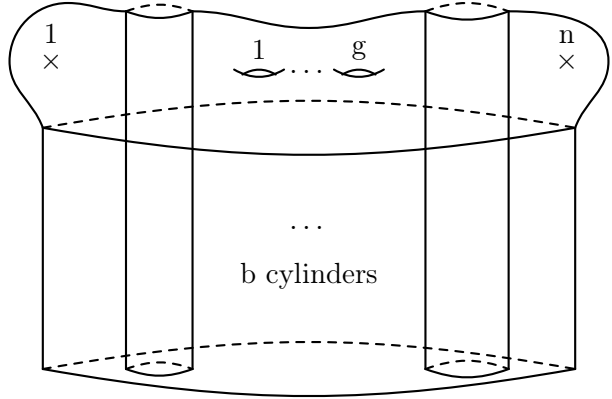

    \centering
    \drawRiemannSurface
    \caption{The geometry of $\Sigma(\Gamma^n_{g,b})$ consists of $b$ boundary cylinders and a connected core Riemann surface $\mathcal{S}(\Gamma^n_{g,b})$. The cylinders are glued to $\mathcal{S}(\Gamma^n_{g,b})$, upon which the on-shell bulk vertex operators are inserted.}
    \label{drawRiemannSurface}
\end{figure}

An arbitrary worldsheet $\Sigma\left(\Gamma^n_{g,b}\right)$ associated with the correlator $\langle \!\langle \mathcal{O}_{V_1}\left(k_1\right) \cdots \mathcal{O}_{V_n}\left(k_n\right) \rangle \!\rangle$ is a Riemann surface of genus $g$ with $b$ boundaries. The defining geometric feature of this worldsheet is that the neighborhoods of its boundaries form semi-infinite cylinders of fixed circumferences. The precise circumference of each cylinder is uniquely determined by the widths of the constituent semi-infinite strips associated with the internal open string propagators. Topologically, these asymptotic cylinders are glued to a core, connected Riemann surface denoted as $\mathcal{S}(\Gamma^n_{g,b})$:
\begin{equation}\label{S}  
\Sigma\left(\Gamma^n_{g,b}\right)=\mathcal{S}(\Gamma^n_{g,b})\cup_{i=1}^b H_i, 
\end{equation}
where $H_{i}$ represents an uncapped, semi-infinite cylinder; see Figure~\ref{drawRiemannSurface}. The on-shell bulk vertex operators are inserted exclusively on the core surface $\mathcal{S}(\Gamma^n_{g,b})$. For instance, in the tree-level disk sector, $\mathcal{S}(\Gamma^4_{0,1})$ corresponds to a disk featuring four internal marked points that host the bulk vertex operator insertions. \par

\begin{figure}[t!]
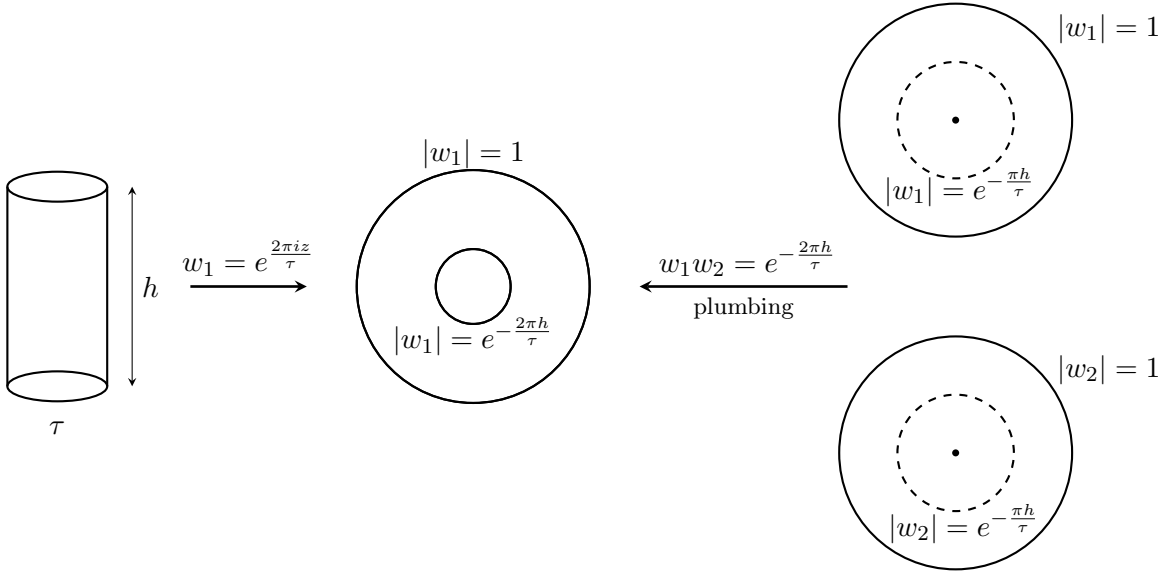

    \centering
    \Cylplumb
    \caption{A cylinder of height $h$ and circumference $\tau$ maps conformally onto an annulus with a unit outer radius and an inner radius of $e^{-2\pi h/\tau}$. This geometry can alternatively be constructed by gluing a pair of punctured unit disks via their local coordinates $w_{1}$ and $w_{2}$, subject to the standard plumbing relation $w_1 w_2 = e^{-2\pi h/\tau}$.}
    \label{Cylplumb}
\end{figure}

Assume that the $i$-th boundary cylinder possesses a circumference $\tau_{i}$, matching the sum of the widths of all its constituent open string strips. Rather than working directly with semi-infinite cylinders, it is computationally convenient to consider cylinders of finite height $h$ and defer the asymptotic limit $h \to \infty$ to the final stage of the analysis. Crucially, a cylinder of height $h$ and circumference $\tau_{i}$ maps conformally onto an annulus with a unit outer radius and an inner radius of $e^{-\frac{2\pi h}{\tau _{i}}}$. Such an annulus is systematically constructed by gluing a pair of punctured unit disks, $D^i_1 = \{\lvert w^i_1 \rvert \leq 1\}$ and $D^i_2 = \{\lvert w^i_2 \rvert \leq 1\}$, by identifying their respective local coordinates via the standard plumbing fixture relation:
\begin{equation}\label{plumb}
w^i_1w^i_2=e^{-\frac{2\pi h}{\tau_i}}.
\end{equation}
As a result, the specific diagrammatic contribution from the graph $\Gamma_{g,b}^{n}$ to the $S$-matrix takes the form:
\begin{align}\label{Wcorr}
&\sum_{\Gamma^n_{g,b}}\int_{0}^{\infty} \prod_{i=1}^{d_{g,n,b}}dt_i({\Gamma^n_{g,b}})  e^{-t_i({\Gamma^n_{g,b}})}\langle{V}_{k_1}^1(z_1)\cdots{V}^n_{k_n}(z_n)\mathcal{B}(\Gamma^n_{g,b})\rangle_{{\Sigma}(\Gamma^n_{g,b})} \nonumber\\
&=\sum_{r_1,\cdots,r_b}\int_{\mathcal{M}_{g,n+b}}\prod_{i=1}^{d_{g,n,b}-b}du_i\left\langle {V}_{k_1}^1(z_1)\cdots{V}^n_{k_n}(z_n) {V}^{r_1}_0(z_{n+1})\cdots V^{r_b}_0(z_{n+b})\mathcal{B}_{g,n+b}\right\rangle_{ \mathcal{R}_{g,n+b}} \nonumber\\
&\qquad \qquad \qquad \qquad \prod_{j=1}^b\left(\lim_{h\to\infty}\int_{0}^{\infty}d\tau_j~\frac{2\pi}{\tau_j}e^{-\frac{2\pi h }{\tau_j} (N_{r_j}-1)}e^{-\tau_j}\left\langle b_0{\hat V}_{0}^{r_j}\right\rangle_{D_2^j}\right).
\end{align}
Here, $N_{r_j}$ denotes the level of the vertex operator. The worldsheet correlator on the right-hand side (R.H.S.) is defined on the genus-$g$ Riemann surface $\mathcal{R}_{g,n+b}$ with $(n+b)$ marked points. This surface is constructed by replacing the $b$ boundary cylinders of $\mathcal{S}(\Gamma^n_{g,b})$ with the local disks $D_1^1, \dots, D_1^b$. The integration domains on the left-hand side (L.H.S.) and the R.H.S. match because the Schwinger parameter domains of all trivalent graphs with $(n+b)$ external legs and genus $g$ provide a single cover of the moduli space $\mathcal{M}_{g,n+b}$ \cite{Giddings:1986wp}. The operator-valued top form on $\mathcal{M}_{g,n+b}$ is built using the $b$-ghost insertion $\mathcal{B}_{g,n+b}$. The factor of $\frac{2\pi}{\tau_i}$ arises via the conformal transformation of the $b$-ghost zero mode from the boundary cylinder to the unit disk. Additionally, the disk one-point function $\langle b_0 {\hat V}_{0}^{r_j}\rangle_{D_2^j}$ of the conjugate vertex operator ${\hat V}_{0}^{r_j}$ is evaluated subject to $D25$-brane boundary conditions. This mapping on the R.H.S. comports naturally with the mechanism for the closure of OSFT worldsheet boundaries proposed in \cite{Gaiotto:2001ji}. \par

In the infinite-height limit ($h \to \infty$), the only states in the summation over $\{r_i\}$ yielding non-vanishing contributions are the zero-momentum closed string tachyon state $\vert T\rangle = c_1\bar c_1\vert 0\rangle$ and the zero-momentum massless state:
\begin{equation}\label{m0state}
\vert D\rangle =\left(\eta_{\mu\nu} c_1\bar c_{-1}\alpha^{\mu}_{-1}\bar \alpha^{\nu}_{-1}+\frac{1}{2}c_1c_{-1}-\frac{1}{2}\bar c_1\bar c_{-1}\right)\vert 0\rangle.
\end{equation}
This fact can be verified by expanding the effective action for the $D25$-brane around the flat background. However, the terms in the expansion for which a tachyon is exchanged along the $i$-th boundary cylinder (of circumference $\tau_i$) are governed by the following seemingly divergent integral:
\begin{equation}\label{integral}
\mathcal{I} = \lim_{h\to\infty}\int_0^{\infty} \frac{d\tau_i}{\tau_i} \, e^{\frac{2\pi h}{\tau_i}} \, e^{-\tau_i}.
\end{equation} 
The divergent behavior of this integral stems directly from applying the Schwinger representation to a tachyonic channel. \par
 
We clarify this behavior by considering the zero-momentum tachyon closed string propagator in the Siegel gauge:
\begin{equation}\label{prop}
\mathcal{P} = -b_{0}^{+}\,b_{0}^{-} \equiv b_{0}^{+}\,b_{0}^{-} \int_{0}^{\infty} ds \, e^{s}.
\end{equation}
The naive Schwinger parameterization yields a divergent integral. This suggests that the divergent integral~\eqref{integral} can be made finite if we can extract the contribution to this integral from the CSFT propagator. Therefore, we shall split the integral on the R.H.S. of~\eqref{split} into finite and divergent pieces:
\begin{equation}\label{split}
\mathcal{I}\left(h\right)= \int_0^{\infty} \frac{d\tau_i}{\tau_i} \, e^{\frac{2\pi h}{\tau_i}} \, e^{-\tau_i} =\mathcal{I}_{\text{fin}}\left(h\right)+\mathcal{I}_{\text{div}}\left(h\right).
\end{equation}
Following the framework in \cite{Sen:2024npu}, we identify the divergent piece,
\begin{equation}\label{iIntegral}
\mathcal{I}_{\text{div}}\left(h\right) = \int_0^{\Lambda} \frac{d\tau_i}{\tau_i} \, e^{\frac{2\pi h}{\tau_i}} \, e^{-\tau_i},
\end{equation}  
as the contribution from an open-closed SFT propagator in the Siegel gauge with the Schwinger parameter denoted as $q_i$. Here, $\Lambda$ is a sufficiently large real number. The Schwinger parameter for the closed string propagator can be related to the parameter $\tau_i$ and the cylinder height $h$ using the following identification:
\begin{equation}\label{schwing}
dq_i~e^{q_i}=\frac{d\tau_i}{\tau_i}~ e^{-\tau_i+\frac{2\pi h}{\tau_i}}.
\end{equation}
This identification allows us to express $\mathcal{I}_{\text{div}}\left(h\right)$ as the following contribution from the closed string propagator:
\begin{equation}\label{Cintegral}
\mathcal{I}_{\text{div}}\left(h\right) = \int_{\frac{2\pi h}{\Lambda}}^{\infty} dq_i \,  e^{q_i},
\end{equation} 
which clearly vanishes in the $h \to \infty$ limit. Interestingly, the finite piece,
\begin{equation}\label{Fintegral}
\mathcal{I}_{\text{fin}}\left(h\right) =\int_{\Lambda}^{\infty} \frac{d\tau_i}{\tau_i} \, e^{\frac{2\pi h}{\tau_i}} \, e^{-\tau_i},
\end{equation} 
also vanishes due to exponential damping. Consequently, the total factor $\mathcal{I}$ vanishes. \par

This results in a drastic simplification of the total contribution to the $n$-point OSFT correlation function from all graphs of a fixed topology:
\begin{align}\label{Wcorrinf}
&\sum_{\Gamma^n_{g,b}}\int_{0}^{\infty} \prod_{j=1}^{d_{g,n,b}}dt_j({\Gamma^n_{g,b}}) \, e^{-t_j({\Gamma^n_{g,b}})}\,\langle{V}_{k_1}^1(z_1)\cdots{V}^n_{k_n}(z_n)\mathcal{B}(\Gamma^n_{g,b})\rangle_{{\Sigma}(\Gamma^n_{g,b})} \nonumber\\
&= \frac{1}{b!}\left(Q\right)^b\int_{\mathcal{M}_{g,n+b}}\prod_{j=1}^{d_{g,n+b}-b}du_j\left\langle {V}_{k_1}^1(z_1)\cdots{V}^n_{k_n}(z_n){ V}^D_0(z_{n+1})\cdots { V}^D_0(z_{n+b})\mathcal{B}_{g,n+b}\right\rangle_{ \mathcal{R}_{g,n+b}},
\end{align}
where $V^D_0$ is the vertex operator corresponding to the state $\vert D\rangle$, $\{u_j\}$ are the coordinates on the moduli space $\mathcal{M}_{g,n+b}$ of the genus-$g$ Riemann surface with $n+b$ punctures, $\mathcal{B}_{g,n+b}$ denotes the associated $b$-ghost insertion, and the factor $Q$ is given by:
\begin{equation}\label{Q}
Q = 2\pi g_o^2\left\langle b_0{\hat V}^D_0\right\rangle_{\text{disk}}\int_{0}^{\infty}\frac{d\tau}{\tau} e^{-\tau}.
\end{equation}
Remarkably, the boundaries of the open string worldsheet are closed dynamically by replacing each boundary cylinder with a disk supporting a zero-momentum dilaton vertex operator insertion. The resulting expression is precisely proportional to the genus-$g$ contribution to a closed string scattering amplitude containing $n+b$ external states, where $b$ of these states are the zero-momentum external states $\vert D\rangle$.

\subsection{The emergence of the  closed string \(S\)-matrix}\label{emcloseds}

Although the final result in the previous subsection is proportional to the genus-$g$ contribution of a closed string amplitude with $n+b$ external states, the proportionality factor remains sensitive to the $D25$-brane boundary conditions of the underlying OSFT. Furthermore, the modular integral contained within this pre-factor is logarithmically divergent. \par

Nevertheless, progress can be made by leveraging the dilaton theorem \cite{Bergman:1994qq} and the soft theorem for zero-momentum gravitons \cite{Sen:2017xjn}. Out of the $n+b$ external states in the closed string amplitude~\eqref{Wcorrinf}, $b$ are the zero-momentum states $\vert D\rangle$, each of which is a linear superposition of the trace of the zero-momentum graviton state and the zero-momentum dilaton state. The physical essence of the dilaton theorem is straightforward: inserting an on-shell, zero-momentum dilaton vertex operator into a string scattering amplitude is equivalent to differentiating that amplitude with respect to the closed string coupling constant. For a genus-$g$ contribution to a scattering amplitude featuring $b$ insertions of $\vert D\rangle$, the soft theorems for the zero-momentum dilaton and graviton states dictate that:
\begin{align}\label{reduction}
\int_{\mathcal{M}_{g,n+b}}&\prod_{j=1}^{d_{g,n,b}-b}du_j\left\langle {V}_{k_1}^1(z_1)\cdots{V}^n_{k_n}(z_n){ V}^D_0(z_{n+1})\cdots { V}^D_0(z_{n+b})\mathcal{B}_{g,n+b}\right\rangle_{ \mathcal{R}_{g,n+b}} \nonumber\\
&= Y^{b}\int_{\mathcal{M}_{g,n}}\prod_{j=1}^{d_{g,n,0}}du_j\left\langle {V}_{k_1}^1(z_1)\cdots{V}^n_{k_n}(z_n)\mathcal{B}_{g,n}\right\rangle_{ \mathcal{R}_{g,n}},
\end{align}
where the differential operator $Y$ is given by:
\begin{equation}\label{Y}
Y = \left(g_s \alpha'^{\frac{24}{2}}\right)^b \left(-\chi_{g,n,0} + \sum_{i=1}^n k_{i}^{\mu}\frac{\partial}{\partial k_i^{\mu}}\right).
\end{equation}
Substituting this structural reduction into~\eqref{Wcorrinf} and summing over all boundary cylinder configurations $b$ yields:
\begin{align}\label{OV}
\langle \!\langle\mathcal{O}_{V_1}\left(k_1\right)  \cdots  \mathcal{O}_{V_n}\left(k_n\right)  \rangle \!\rangle&=\sum_{g=0}^{\infty}\sum_{b=1}^{\infty} \frac{\left(QY \right)^{b} }{b!}\int_{\mathcal{M}_{g,n}}\prod_{j=1}^{d_{g,n,0}}du_j\left\langle {V}_{k_1}^1(z_1)\cdots{V}^n_{k_n}(z_n)\mathcal{B}_{g,n}\right\rangle_{ \mathcal{R}_{g,n}} \nonumber\\
&=\sum_{g=0}^{\infty}\left(e^{QY} - 1\right)\int_{\mathcal{M}_{g,n}}\prod_{j=1}^{d_{g,n,0}}du_j\left\langle {V}_{k_1}^1(z_1)\cdots{V}^n_{k_n}(z_n)\mathcal{B}_{g,n}\right\rangle_{ \mathcal{R}_{g,n}}.
\end{align} 
Interestingly, the quantity $Q$ defined in~\eqref{Q} is negative and logarithmically divergent. The negative sign of $Q$ originates from the disk one-point function of the dilaton, whose regularized value is given by:
\begin{equation}\label{dilaton}
\langle b_0 {\hat V}_0^{D}(0)\rangle _{\text{disk}} = -\frac{2 T_{25}}{g_c},
\end{equation}
where $T_{25}$ represents the physical tension of the unstable $D25$-brane. Because $Q$ possesses a large, negative divergent real part, the exponential factor $e^{QY}$ on the right-hand side (R.H.S.) of~\eqref{OV} vanishes in the strict cutoff limit. Suppressing this vanishing exponential contribution, we obtain:
\begin{align}\label{OVV}
\langle \!\langle\mathcal{O}_{V_1}\left(k_1\right)  \cdots  \mathcal{O}_{V_n}\left(k_n\right)  \rangle \!\rangle=-\sum_{g=0}^{\infty}g_o^{-2\chi_{g,n,0}}\int_{\mathcal{M}_{g,n}}\prod_{j=1}^{d_{g,n,0}}du_j\left\langle {V}_{k_1}^1(z_1)\cdots{V}^n_{k_n}(z_n)\mathcal{B}_{g,n}\right\rangle_{ \mathcal{R}_{g,n}}.
\end{align}
We have thus derived the formal genus expansion of the bosonic closed string $S$-matrix in 26-dimensional flat spacetime—as traditionally defined via Polyakov path integrals \cite{Polyakov:1981rd}—purely from the correlation functions of OSFT gauge-invariant operators evaluated around the tachyon vacuum background, up to an overall minus sign.\footnote{The issue of an overall minus sign can be resolved through a detailed comparative analysis of the closed string measure and the modular measure naturally constructed by the OSFT geometric patches.} However, because the worldsheet bulk vertex operators used to assemble these gauge-invariant operators were assumed to satisfy strict tree-level on-shell conditions from the outset, this direct mapping correctly defines the physical $S$-matrix only at the tree level. \par

This subtle restriction underscores why the all-orders genus expansion of the bosonic closed string $S$-matrix remains structurally ill-defined and plagued with ultraviolet loop divergences when constructed naively from first-quantized closed string theory. To build an $S$-matrix that remains well-defined and unitary to all orders in perturbation theory, one must implement a fully off-shell Lehmann–Symanzik–Zimmermann (LSZ) reduction formula \cite{Lehmann}. This requires evaluating the correlation functions of OSFT operators constructed from off-shell worldsheet bulk operators, followed by a systematic application of the LSZ prescription. Such an extension is anticipated to systematically reproduce the complete, ultraviolet- and infrared-finite genus expansion of the bosonic closed string $S$-matrix described in \cite{Pius:2013sca, Pius:2014iaa, Sen:2014pia}.

\section{Discussion}\label{sec:Discussion}

We have demonstrated that the OSFT correlation functions of gauge-invariant operators, when evaluated perturbatively around the tachyon vacuum, systematically compute the closed string \(S\)-matrix for the associated on-shell states. This result directly implies that quantum gravity emerges dynamically from the open string tachyon vacuum background. Consequently, our findings establish Witten’s cubic OSFT as a fully calculable, non-perturbative formulation for quantum gravity in asymptotically flat spacetimes. \par

While this work focuses exclusively on the unstable \(D25\)-brane, extending this analysis to unstable \(Dp\)-branes for \(p<25\) is a natural and crucial next step. Based on Sen's conjecture, one expects to reproduce the same results around the tachyon vacuum. The OSFT defined on a \(Dp\)-brane explicitly breaks Lorentz invariance along the \(25-p\) transverse spacetime directions. Therefore, this extension will be instrumental in uncovering the mechanism that recovers full Lorentz invariance in the tachyon vacuum. Similarly, it would be interesting to extend the bosonic string field theory analysis presented here to open superstring field theory. Such an extension is anticipated to reproduce the exact \(S\)-matrix of closed superstring theory in 10-dimensional flat spacetime. \par

Another vital direction involves investigating closed string tachyon condensation. Although a well-defined action for bosonic closed string field theory exists, analyzing this phenomenon has remained notoriously difficult due to the highly non-polynomial nature of the action and the immense challenges inherent in its explicit construction \cite{Yang:2005rx}. By offering an alternative, fully calculable framework for closed string dynamics, the machinery developed in this paper can be utilized to compute the closed string tachyon potential and systematically analyze its rolling behavior. \par

The construction of worldsheets associated with the correlation functions of gauge-invariant operators via perturbation theory around the tachyon vacuum bears a close resemblance to the gauge/string duality framework formulated in terms of gluing Feynman diagrams \cite{Gopakumar:2003ns,Gopakumar:2004qb,Gopakumar:2005fx,David:2006qc}. It would be interesting to explore this connection in closer detail. \par

A more ambitious prospect is to investigate whether the five tachyon-free closed superstring theories can be realized as perturbative expansions around specific tachyon-free stable coherent states of open bosonic string field theory constructed using the gauge invariant operators. This might initially seem implausible, given that the bosonic string conventionally yields only spacetime bosons, whereas superstring theories incorporate spacetime fermions. However, bosonic closed strings compactified to 10 dimensions on the \(SO(32)\) lattice are known to possess a rich spectrum of non-perturbative fermionic excitations. It is natural to attempt to verify the conjecture that these fermionic states become dynamically lighter during closed string tachyon condensation \cite{David:2001nq}. If closed superstring field theories can indeed be established as fluctuations around these stable, tachyon-free coherent states, Witten’s cubic open string field theory may offer a novel, unified pathway toward uncovering the non-perturbative structure of M-theory.


\acknowledgments

It is a pleasure to thank Ashoke Sen for insightful discussions and crucial suggestions. We are also grateful to Sujay Ashok, Rajesh Gopakumar, Alok Laddha, R. Loganayagam, Gautam Mandal, Sameer Murthy, Venkata Suryanarayana Nemani, Ebin Paul and Spenta Wadia for their valuable comments and feedback.



\end{document}